\documentclass[10pt]{iopart}
\pdfoutput=1

\expandafter\let\csname equation*\endcsname=\relax
\expandafter\let\csname endequation*\endcsname=\relax
\usepackage{color}
\usepackage{amsmath}
\usepackage{amssymb}
\usepackage{enumerate}
\usepackage{graphicx}
\usepackage{mathbbol}
\usepackage{amsfonts}
\usepackage{url}

\newcommand{\nn}{\nonumber}

\newcommand{\Prob}{{\rm Prob}}
\newcommand{\bea}{\begin{eqnarray}}
\newcommand{\eea}{\end{eqnarray}}
\newcommand{\beq}{\begin{equation}}
\newcommand{\eeq}{\end{equation}}

\def\XXint#1#2#3{{\setbox0=\hbox{$#1{#2#3}{\int}$}
 \vcenter{\hbox{$#2#3$}}\kern-.5\wd0}}

\usepackage[utf8]{inputenc}
\usepackage{amsmath}
\usepackage{hyperref}
\usepackage{graphicx}
\usepackage{amsfonts}
\usepackage{amsthm}
\usepackage{cases}
\usepackage{bm}
\usepackage{amssymb}

\usepackage{color}
\definecolor{Blue}{rgb}{0.00, 0.00, 1.00}
\definecolor{Red}{rgb}{1.00, 0.00, 0.00}

\hypersetup{
    colorlinks=true,       
    linkcolor=red,          
    citecolor=blue,        
    filecolor=magenta,      
    urlcolor=cyan           
}

\newcommand{\be}{\begin{equation}}
\newcommand{\ee}{\end{equation}}

\newcommand{\beqn}{\begin{eqnarray}}
\newcommand{\eeqn}{\end{eqnarray}}

\DeclareMathOperator{\sgn}{sgn}

\DeclareMathOperator{\I}{I}

\DeclareMathOperator{\erf}{erf}

\newcommand{\pFq}[5]{{}_{#1}\mathrm{F}_{#2} \left( \begin{array}{c} #3
\\ #4 \end{array} ; #5 \right)}

\makeatletter
\renewcommand\@appendixstar{\@@par
 \ifnumbysec 
 \@addtoreset{table}{section}
 \@addtoreset{figure}{section}\fi
 \setcounter{section}{0}
 \setcounter{subsection}{0}
 \setcounter{subsubsection}{0}
 \setcounter{equation}{0}
 \setcounter{figure}{0}
 \setcounter{table}{0}
 \def\thesection{\Alph{section}} 
 \def\theequation{\ifnumbysec
      \Alph{section}.\arabic{equation}\else
      \Alph{section}\arabic{equation}\fi}
 \def\thetable{\ifnumbysec
      \Alph{section}\arabic{table}\else
      A\arabic{table}\fi}
 \def\thefigure{\ifnumbysec
      \Alph{section}\arabic{figure}\else
      A\arabic{figure}\fi}}
\makeatother
\begin{document}
\title[]{Universal survival probability for a correlated random walk and applications to records}

\author{Bertrand Lacroix-A-Chez-Toine}
\address{Department of Physics of Complex Systems,
 Weizmann Institute of Science,
 7610001 Rehovot, Israel}

\author{Francesco Mori}
\address{LPTMS,  CNRS,  Univ.   Paris-Sud,  Universit\'e  Paris-Saclay,  91405  Orsay,  France}

\begin{abstract}

We consider a model of space-continuous one-dimensional random walk with simple correlation between the steps: the probability that two consecutive steps have same sign is $q$ with $0\leq q\leq 1$. The parameter $q$ allows thus to control the persistence of the random walk. We compute analytically the survival probability of a walk of $n$ steps, showing that it is independent of the jump distribution for any finite $n$. This universality is a consequence of the Sparre-Andersen theorem for random walks with uncorrelated and symmetric steps. We then apply this result to derive the distribution of the step at which the random walk reaches its maximum and the record statistics of the walk, which show the same universality. In particular, we show that the distribution of the number of records for a walk of $n\gg 1$ steps is the same as for a random walk with $n_{\rm eff}(q)=n/(2(1-q))$ uncorrelated and symmetrically distributed steps. We also show that in the regime where $n\to \infty$ and $q\to 1$ with $y=n(1-q)$, this model converges to the run-and-tumble particle, a persistent random walk often used to model the motion of bacteria. Our theoretical results are confirmed by numerical simulations.
\end{abstract}

\maketitle


\section{Introduction and presentation of the model}

\subsection{Introduction}

Records play a central role in a number of different contexts ranging from climate, sports to finance. It is thus essential to be able to predict the statistics of both the occurrence and the increments of these records. They also have interesting applications in the physics of spin glasses \cite{SRK06,S07,LDW09} or superconductors \cite{OJNS05} (see \cite{W13,GMS17} for reviews). The classical theory of records was developed around independent and identically distributed random variables \cite{ABN98,N04}. Many refinements have been considered ever since with e.g. independent but non-identically distributed random variables (see \cite{W13,SM14} and reference therein). While these statistics are well understood for systems whose time evolution is uncorrelated, it is in general very difficult to obtain analytical results for any model with  strong correlations in time. 

\begin{figure}[h]
\centering
\includegraphics[width=0.6\textwidth]{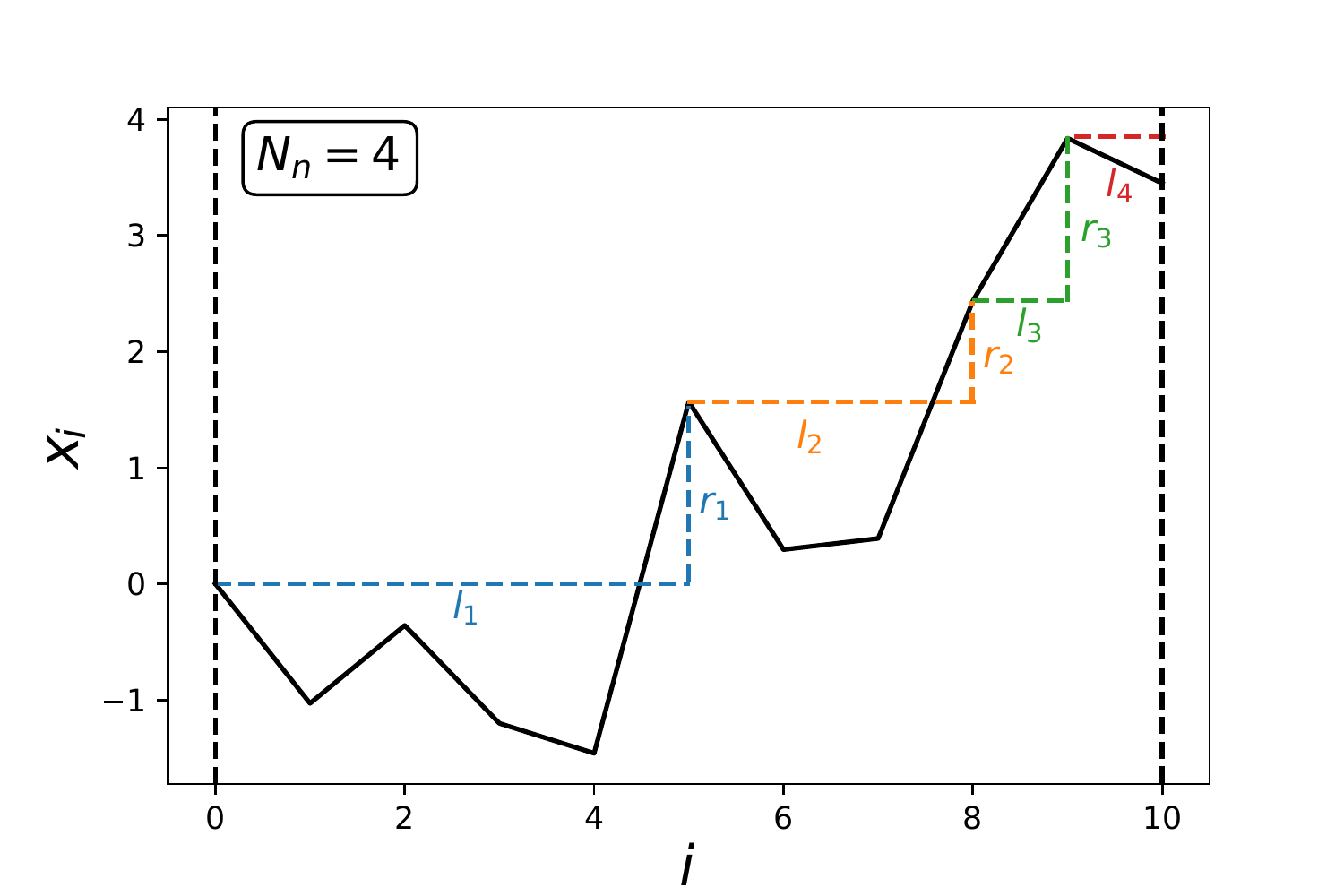}
\caption{Sketch of a random walk of $n=10$ steps with $N_n=4$ records (the initial position is a record) with $l_i$ the age of the record $i$, i.e. the times for which it is standing, and $r_i$ the increment associated with record $i$, i.e. the difference between the value of record $i+1$ and $i$.}\label{Fig_sketch_rec}
\end{figure}

The statistics of the number $N_n$ of records up to step $n$ and their ages $l_i$'s for $i=1,\cdots,N_n$, i.e. the time during which a record is standing (see Fig. \ref{Fig_sketch_rec}), were computed exactly for the first time for a random walk in \cite{MZ08}. It was shown that these statistics are {\it universal} for any number of steps $n$, that is they are independent of the distribution of the steps of the random walk, as long as it is continuous and symmetric. The distribution of the record increments $r_j$'s for $j=1,\cdots,N_n-1$ (see Fig. \ref{Fig_sketch_rec}) were computed in \cite{GMS16} and found on the contrary to depend explicitly on the distribution of the steps. 

To compute these probabilities, a key building block turns out to be the survival probability of the random walk \cite{GMS17,SM14,MZ08}. Survival (or persistence) probability, i.e. the probability that an observable of a stochastic system does not change sign up to a given number of steps (or time), and the associated first-passage probability, have been themselves the subject of intense theoretical and experimental studies since the 50s (see \cite{M99,R01,BMS13,AS15} for reviews on the subject). This probability is often highly non-trivial to compute as it depends on the whole history of the system. The survival probability for random walks is one of the first central result on the subject and was derived by Sparre-Andersen \cite{SA55}. It was proven that for any symmetric space-continuous random walk, starting from the origin, it takes the universal value
\be
S_n(0)=\Prob\left[x_0\geq 0,x_1\geq 0,\cdots,x_n\geq 0|x_0=0\right]={{2n}\choose{n}}2^{-2n}\;.\label{SA}
\ee 
The survival probability for a random walk starting from an arbitrary position $x$ can simply be related to the extreme value statistics of the process (see \cite{M10} for a simple illustration) and is expressed exactly via the Pollaczek-Spitzer formula \cite{P52,S56,S57}. However, in this case the universality is broken for finite $n$ and only recovered, although in several universality classes, in the large $n$ limit \cite{MMS17}. This survival probability has been generalised to non-symmetric \cite{SA55} or discrete random walks \cite{MMS20}. 

Since the first results for the record statistics of random walks, several extensions have been explored, including the presence of a drift \cite{MSW12}, for a discrete \cite{MMS20}, a time-continuous \cite{S11}, or multiple walks \cite{WMS12} (see \cite{W13,GMS17} for reviews on the subject). In this article, we obtain a non-trivial result for the survival probability of random walks whose steps are correlated and apply this result to derive the record statistics. We prove that, akin to the Sparre Andersen therorem, these results are universal with respect to the step distribution as long as it is continuous.

The paper is organised as follow. In section \ref{sec_mod}, we present our model of time-correlated random walks. In section \ref{sec_main_res}, we detail our main results on the survival probability and its application to the extreme value and record statistics of the correlated random walk. In section \ref{sec_surv}, we derive the results for the survival probability. In section \ref{sec_max}, we apply this result to compute the distribution of the step at which the random walk reaches its maximum. In section \ref{sec_rec}, we apply our result on the survival probability to compute the statistics of the number of records in the random walk. Finally, in section \ref{sec_conc}, we briefly conclude and present future directions. 

\subsection{Model}\label{sec_mod}

In this article, we consider a one-dimensional random walk with correlated steps defined as 
\be
x_{n+1}=x_n+\sigma_{n}\eta_{n}\;,\label{RW}
\ee
where the random variable $\sigma_{n}=\pm 1$ are binary random variables with the following stochastic evolution
\be
\sigma_n=\begin{cases}
\displaystyle \sigma_{n-1}&\;,\;\;{\rm with\;probability}\;q\;,\\
\displaystyle -\sigma_{n-1}&\;,\;\;{\rm with\;probability}\;1-q\;.
\end{cases}\label{sigma_def}
\ee
The step lengths $\eta_n$'s in Eq. \eqref{RW} are positive i.i.d. random variables drawn from the continuous probability distribution function (PDF) $p(\eta)$. In the following, the term of {\it universality} will always refer with respect to this distribution $p(\eta)$. The process $(\sigma_n,x_n)$ is Markovian. The random walk starts at position $x_0$ (we will often consider $x_0=0$) in state $\sigma_0=\pm $. Note that the record statistics for a model of discrete random walk with correlated steps was recently considered in \cite{K20}. The time correlations in this model where however much stronger as the probability of a positive (resp. negative) step depended on the whole history of the walk and not only on the sign of the last step.
Before presenting the main results, let us mention three special values of $q$ for which this random walk is connected to known processes:
\begin{itemize}
\item For $q=1/2$, the sign of the steps of the random walk are uncorrelated. The random walk is symmetric, the distribution of its steps reading 
\be
p_{\rm sym}(\eta)=\begin{cases}
\displaystyle\frac{p(-\eta)}{2}&\;,\;\;\eta<0\;,\\
&\\
\displaystyle\frac{p(\eta)}{2}&\;,\;\;\eta\geq 0\;.
\end{cases}\label{p_sym}
\ee
\item For $q=0$, the sign of the steps alternates at each iteration. The different positions of the random walker can be interpreted as the positions of a run-and-tumble particle (RTP) at successive tumbling events as seen in Fig. \ref{Fig_RTP_to_RW}. This model, which has been studied extensively recently \cite{TC08,B08,C12,SEB16,Metal18,LDMS19,SK19,DKMSS19,ALW19,HLRM19,DD19,DKD19}, is a time-continuous {\it persistent random walk}, where the position $x(t)$ of the particle at time $t$ follows the Langevin equation
\be
\dot x(t)=v_0 \sigma(t)\;,\label{RTP_lang}
\ee
where $v_0$ is a fixed velocity and $\sigma(t)$ is a telegraphic noise of rate $\gamma$. In the mapping to our model, the usual exponential distribution of the tumbling time $p_{\rm tumbling}(\tau)=\gamma e^{-\gamma \tau}$ of the telegraphic noise is replaced by the distribution $p(\tau)$ of the step's length with the identification $\tau=\eta/v_0$.  

\begin{figure}[h]
\centering
\includegraphics[width=0.47\textwidth]{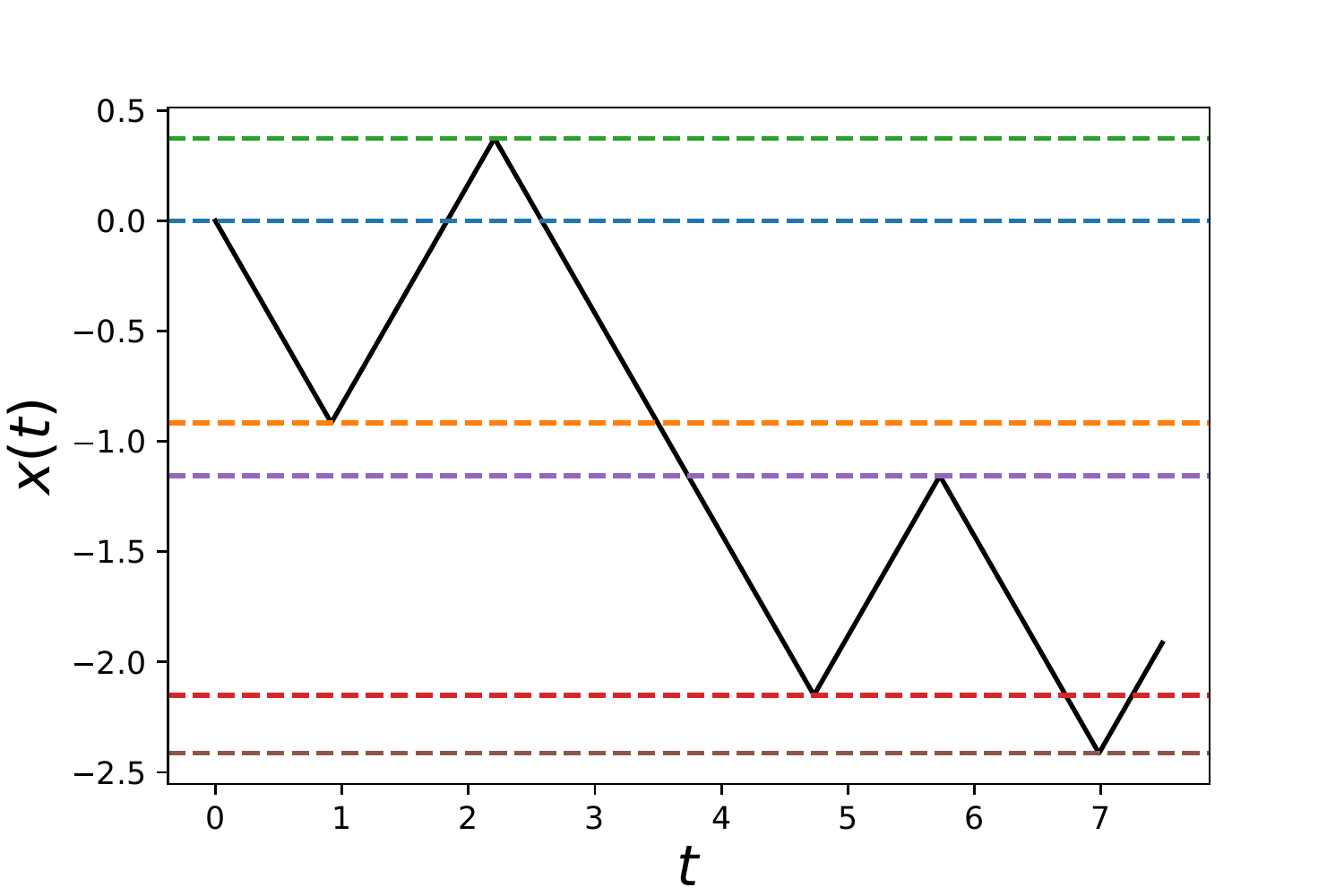}
\includegraphics[width=0.47\textwidth]{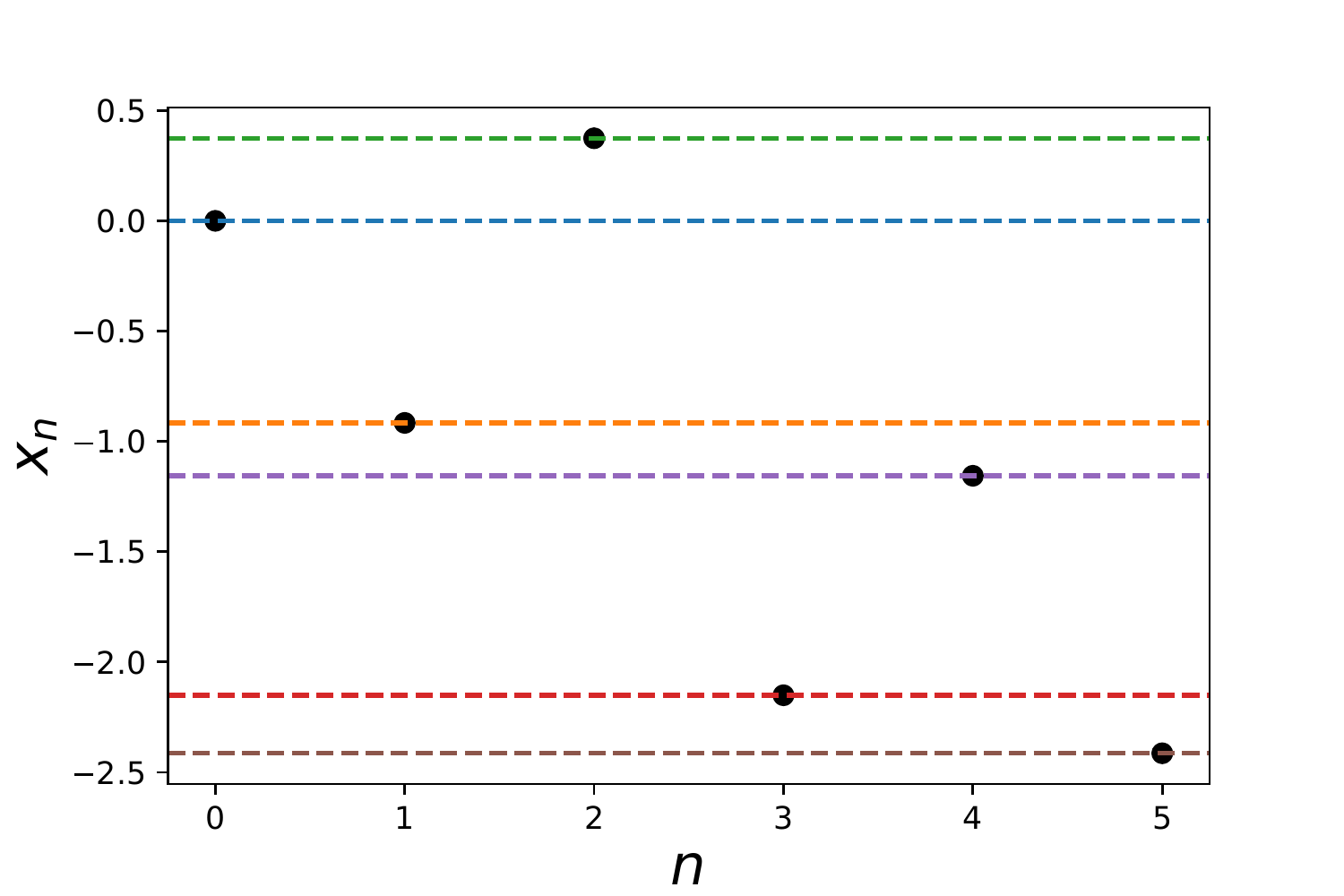}
\caption{Left: Plot of a trajectory $x(t)$ of the RTP at time $t$. Right: Random walk associated to the successive positions $x_n$ of the RTP at the $n^{\rm th}$ tumbling event.}\label{Fig_RTP_to_RW}
\end{figure}

\item For $q=1$ exactly, all the steps have the same sign as $\sigma_0$. This case corresponds to an uncorrelated uni-directional random walk with step distribution $p(\eta)$. Its survival, record and extreme value statistics are trivial. 

In the scaling limit where $n\to \infty$ and $q\to 1$ with $y=n(1-q)=O(1)$ instead, the random walk converges to a model of RTP in continuous time with the correspondence $y=n(1-q)=\gamma t$. Indeed for the RTP, the probability to switch direction during time $dt\ll 1$ is $\gamma dt$. The corresponding probability for the random walk is $1-q$. The persistence probability, i.e. the probability to stay in the same direction for $n$ steps for the random walk is the same as the persistence probability for the RTP during time $t=n dt$. It reads in this limit
\be
P_{\rm per}(t)=q^n\approx e^{-y}=e^{-\gamma t}\;.
\ee

\end{itemize}

Coming back to the case of general $q$, for the random walk defined in Eqs. \eqref{RW} and \eqref{sigma_def}, we will compute the survival probability
\be
S^{+}_n(x;q)=\Prob\left[x_0\geq 0,\;x_1\geq 0,\;x_2\geq 0,\;\cdots,\;x_n\geq 0\;|\;x_0=x,\;\sigma_0=\sigma\right]\;.\label{surv}
\ee
We will put strong emphasis on the case where the random walk starts from the origin $x_0=x=0$ which has interesting applications for the extreme value and record statistics of the random walk. As a first application of this result, we will compute the probability $P_{k,n}^{\max}(q)=\Prob\left[n_{\max}=k\right]$ that the maximum is reached at step $k$ for a random walk of $n$ steps. Finally, we will show that this result allows to compute the probability $R_{m,n}(q)=\Prob\left[N_{n}=m\right]$ that the random walk of $n$ steps has a number $N_n=m$ of records.

\section{Main results}\label{sec_main_res}

\subsection{Survival probability}
\begin{figure}[h]
\centering
\includegraphics[width=0.7\textwidth]{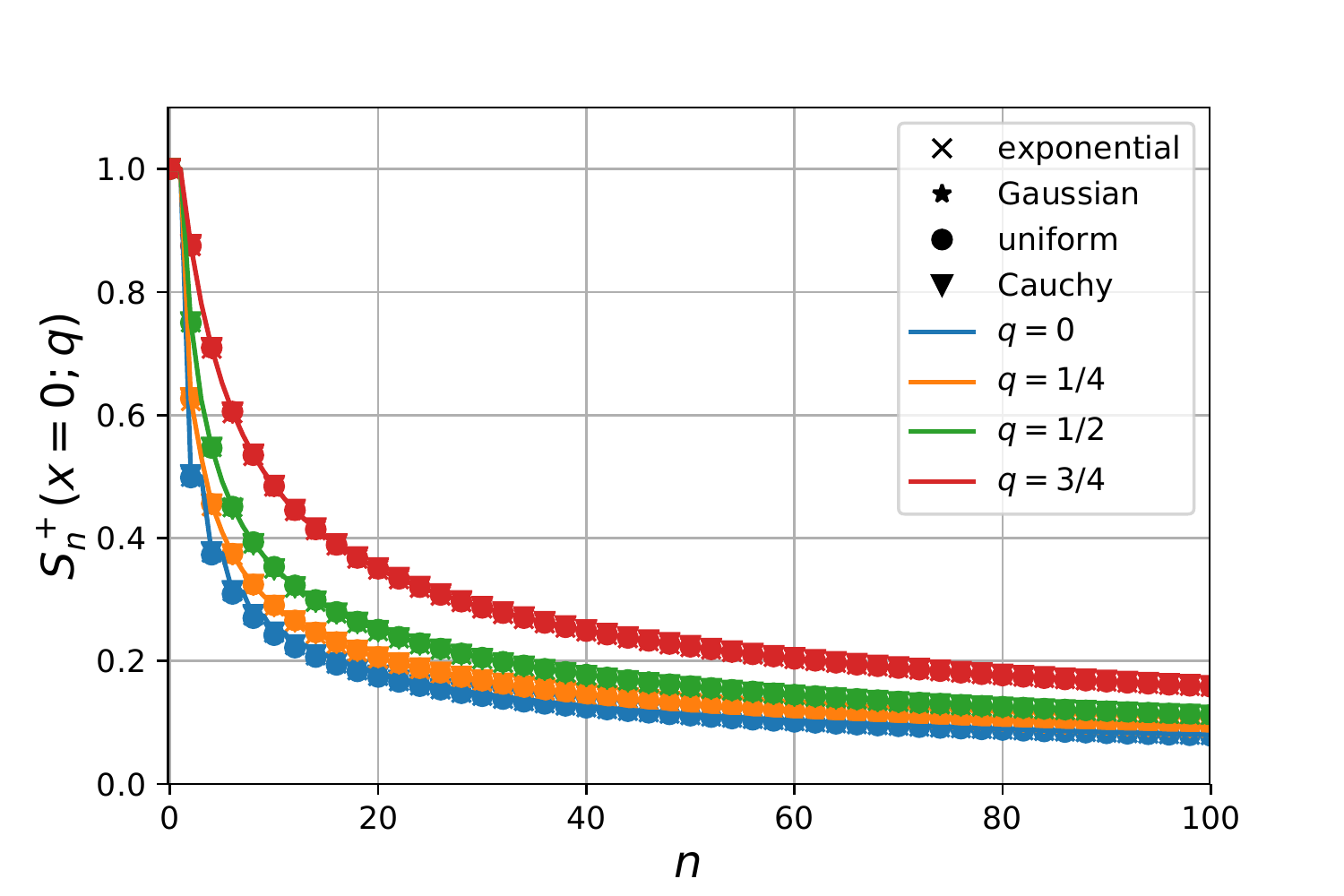}
\includegraphics[width=0.7\textwidth]{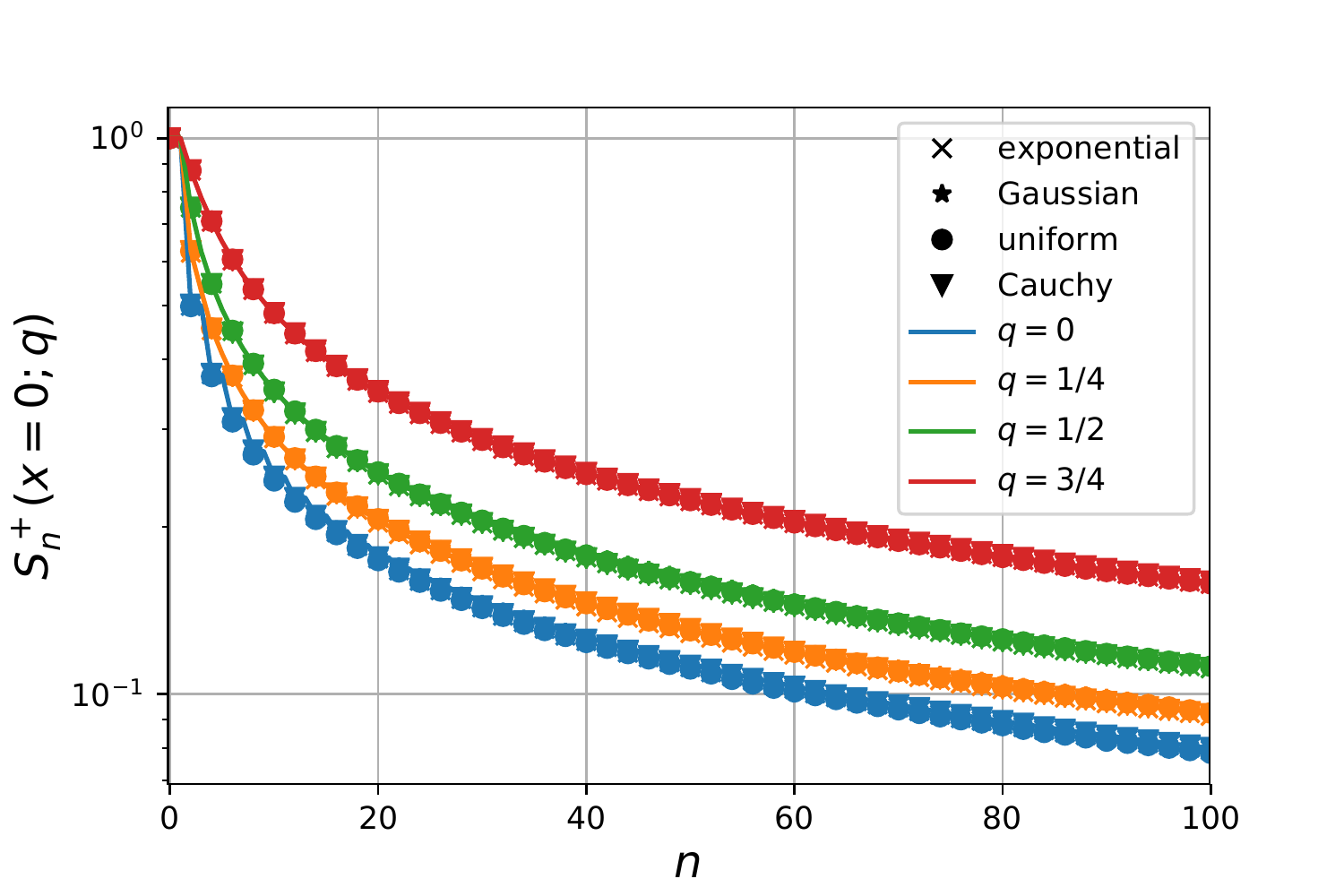}
\caption{Plot of the survival probability $S_n^{+}(x=0;q)$ for different values of $q=0,1/4,1/2,3/4$, respectively in blue, orange, green and red as a function of $n=0,\cdots,100$ in linear (upper panel) and logarithmic (lower panel) scale. The numerical data is obtained by simulating $N=10^5$ runs of the random walk, starting with a positive first step and for several PDF $p(\eta)$ of the steps' lengths. The numerical data collapses exactly for all the different distribution to the analytical prediction given by Eq. \eqref{surv_res}.}\label{Fig_surv_prob}
\end{figure}

The central result of this paper is the universality of the survival probability in the case of a random walk starting from position $x_0=x=0$. Indeed, we show that for any distribution of the steps' lengths $p(\eta)$,
\be
S_n^{+}(x=0;q)=\frac{2^{-2n}}{1-q}{2n\choose n}\pFq{2}{1}{-\frac{1}{2},-n }{\frac{1}{2}-n}{2q-1}\;,\;\;n\geq 1\;,\label{surv_res}
\ee
where $\pFq{2}{1}{a,b }{c}{x}$ is the hypergeometric function defined as
\be
\pFq{2}{1}{a,b }{c}{x}=\sum_{n\geq 0}\frac{(a)_n (b)_n}{n! (c)_n}x^n\;,
\ee
and $(a)_n=\Gamma(a+n)/\Gamma(a)$ the rising factorial (or Pochhammer symbol). Notably, the result in Eq. (\ref{surv_res}) is exact for any finite value of $n$ and for any $0\leq q \leq 1$. As observed in Fig. \ref{Fig_surv_prob}, our theoretical result in Eq. (\ref{surv_res}) is in perfect agreement, for different values of $q$, with numerical simulations performed with different step distributions $p(\eta)$.

Setting $q=1/2$ in Eq. \eqref{surv_res}, one recovers, as expected, the result from the Sparre Andersen theorem $S_n^+(0)=S_n^+(0;1/2)={2n\choose n}2^{-2n+1}$ for $n\geq 1$ (we remind that we have conditioned on the positivity of the first step). Starting from the origin with a negative first step, it is trivial to obtain that for any continuous distribution of the steps' lengths $p(\eta)$ the survival probability is simply $S_n^{-}(x=0;q)=\delta_{n,0}$. In the large $n$ limit, the survival probability asymptotically vanishes as
\be
S_n^{+}(x=0;q)\approx \sqrt{\frac{2}{n\pi(1-q)}}+O(n^{-3/2})\;.\label{S_n_as}
\ee 

We also show that in the scaling regime $n\to \infty$, $q\to 1$ with $y=n(1-q)=O(1)$, the survival probability is a scaling function of the variable $y$,
\be\label{scal_surv_prob_res}
S_n^{+}(x=0;q)\approx {\cal S}(n(1-q))\;,\;\;{\rm with}\;\;{\cal S}(y)=e^{-y}\left(\I_0(y)+\I_1(y)\right)\;.
\ee
We compare in Fig. \ref{Fig_S_y} our analytical prediction in this scaling limit with numerical simulation of the random walk for several distributions of the jump's length $p(\eta)$, showing excellent agreement. The scaling function ${\cal S}(\gamma t)$ is exactly the survival probability up to time $t$ for the run-and-tumble particle (RTP) defined above (see Eq. \eqref{RTP_lang}), starting from $x=0$ with positive speed ($\sigma(0)=+$) \cite{LDMS19}. Note that in the opposite limit $q=0$, where the random walk is mapped exactly on the positions of the RTP at each tumbling event, this scaling function is recovered by considering a waiting time between the steps distributed according to $p_{\rm tumbling}(\tau)=\gamma e^{-\gamma \tau}$.

\subsection{First application: index the maximum}

\begin{figure}[h]
\centering
\includegraphics[width=0.7\textwidth]{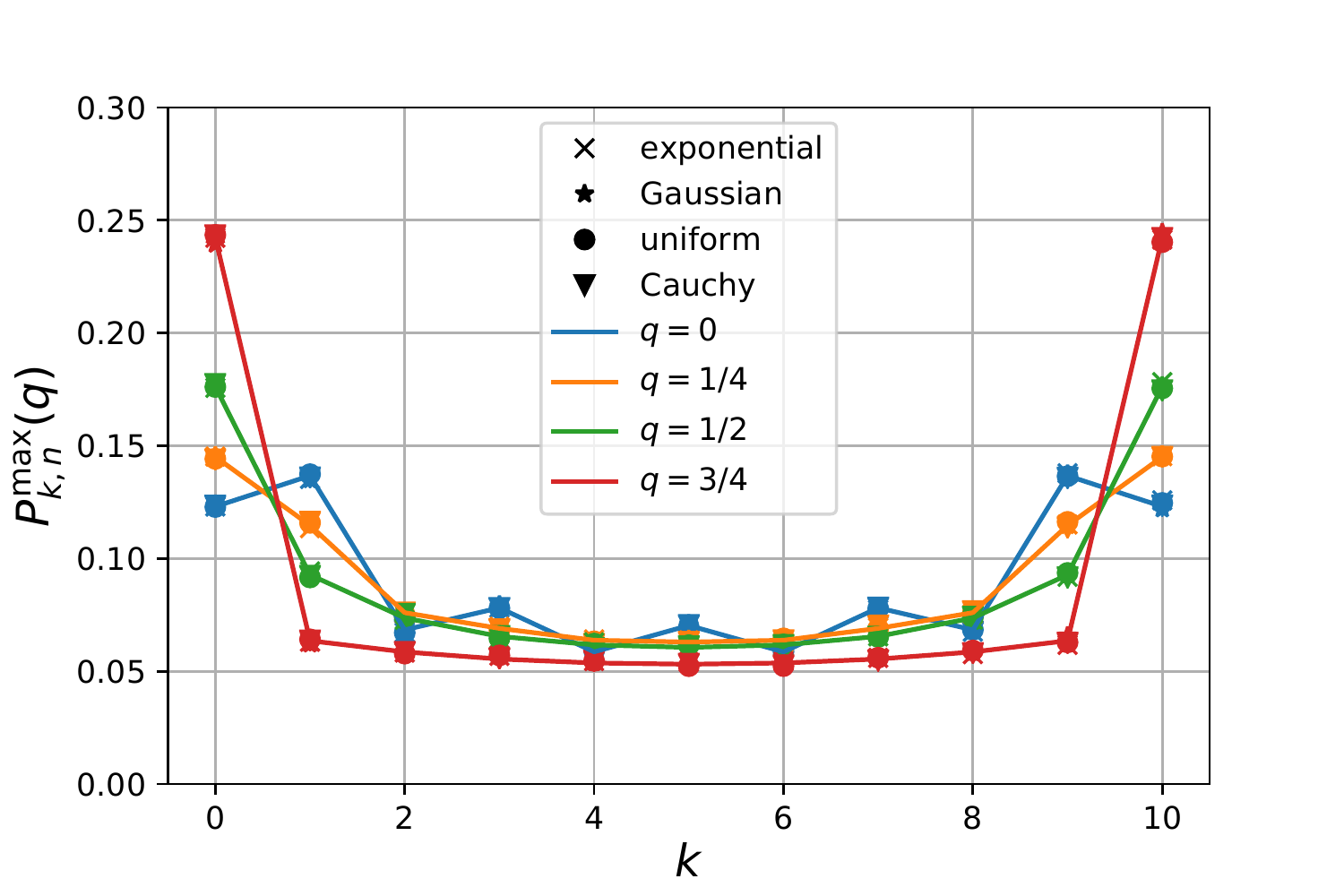}
\caption{Plot of the probability $P_{k,n}^{\max}(q)$ of the index $n_{\max}$ of the maximum for different values of $q=0,1/4,1/2,3/4$, respectively in blue, orange, green and red as a function of $n=0,\cdots,10$. The numerical data is obtained by simulating $N=10^5$ runs of random walks of $n=10$ steps starting with a positive or negative first step with probability $1/2$, for several different PDF $p(\eta)$ of the steps' lengths. The numerical data collapses exactly on the analytical prediction in Eq. \eqref{P_max_n}.}\label{Fig_T_max}
\end{figure}

From our expression for the survival probability starting from $x=0$, we may obtain the distribution of the index $n_{\max}$ of the step at which the maximum of the random walk is reached, i.e. $x_{n_{\max}}=x_{\max}=\max_{k} x_k$. For a random walk with equal probability $1/2$ to start with a positive or negative first step, it reads for $n\geq 1$
\be
P_{k,n}^{\max}(q)=\Prob\left[n_{\max}=k\right]=\begin{cases}
\displaystyle \frac12 S_{n}^{+}(0;q)&\;,\;\;k=0,n\;,\\
&\\
\displaystyle \frac{1-q}{2}S_{k}^{+}(0;q)S_{n-k}^+(0;q)&\;,\;\;0<k<n\;.
\end{cases}
\label{P_max_n}
\ee
In Fig. \ref{Fig_T_max}, we compare our analytical prediction for the distribution of $n_{\max}$ with numerical simulation of the random walk for different step's length distribution $p(\eta)$ and different values of $q$. The collapse on our prediction is excellent and holds for finite $n$.

In the limit where $n\to \infty$ and for any value $0\leq  q<1$ the distribution converges, independently of $q$, to the universal arcsine law
\be
P_{k,n}^{\max}(q)\approx \frac{1}{n}{\cal T}\left(\frac{k}{n}\right)\;,\;\;{\rm with}\;\;{\cal T}(\tau)=\frac{1}{\pi\sqrt{\tau(1-\tau)}}\;.\label{arcsine}
\ee
In Fig. \ref{Fig_T_max_large}, we compare the distribution of $n_{\max}$ obtained via numerical simulation of the process to the arcsine law. In the large $n$ limit, the data collapse for any $q$ and any distribution $p(\eta)$ of the jump's length to our analytical prediction in Eq. \eqref{arcsine}.

On the other hand, in the scaling regime $n\to \infty$ and $q\to 1$ with $y=n(1-q)=O(1)$, the probability of $n_{\max}$ converges to the distribution of the time at which the run-and-tumble particle reaches its maximum \cite{SK19}
\begin{align}
&P_{k,n}^{\max}(q)\approx \frac{1}{n}{\cal P}\left(\frac{k}{n};n(1-q)\right)\;,\\
&{\cal P}(\tau;y)=\frac{y}{2}{\cal S}(y\tau){\cal S}(y(1-\tau))+\frac{\delta(1-\tau)+\delta(\tau)}{2}{\cal S}(y)\;,\label{P_max_RTP}
\end{align}
where $\delta(x)$ is the Dirac delta function and ${\cal S}(y)$ is defined in Eq. \eqref{scal_surv_prob_res}. In Fig. \ref{Fig_P_scal}, we compare the distribution of $n_{\max}$ obtained via numerical simulation of the process in the scaling limit $n\gg 1$ and $1-q\ll 1$ with $n(1-q)=y=O(1)$ and for different distribution $p(\eta)$ of the jumps' lengths, to our analytical prediction in Eq. \eqref{P_max_RTP}, showing excellent agreement.

\subsection{Second application: number of records}

\begin{figure}[h]
\centering
\includegraphics[width=0.7\textwidth]{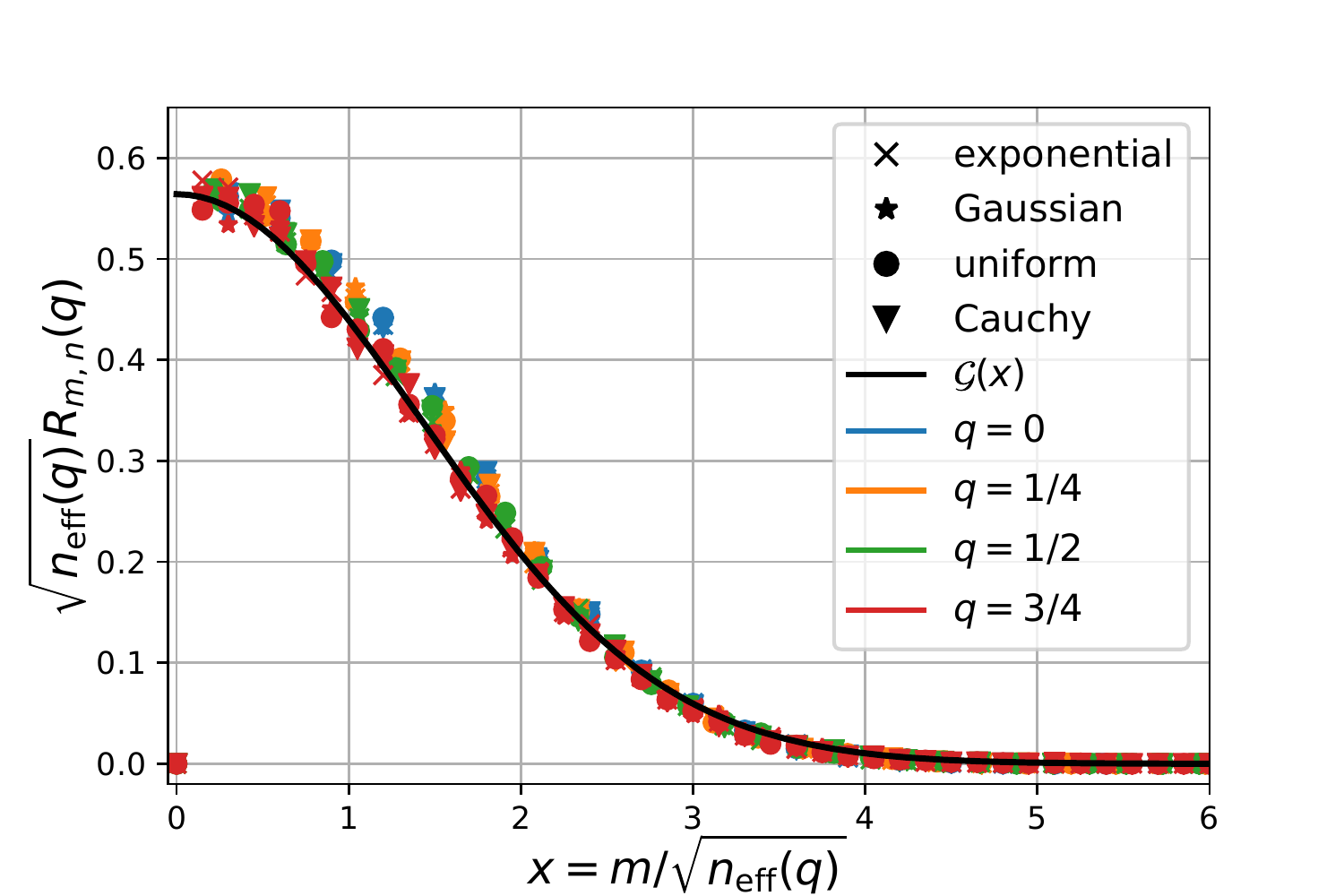}
\caption{Plot of the rescaled probability $\sqrt{n_{\rm eff}(q)}R_{m,n}(q)$ of the number $N_{n}$ of records, with $n_{\rm eff}(q)=n/(2(1-q))$ for different values of $q=0,1/4,1/2,3/4$, respectively in blue, orange, green and red as a function of $x=m/\sqrt{n_{\rm eff}(q)}$. The numerical data is obtained by simulating $N=10^5$ runs of random walks of $n=200$ steps starting with a positive or a negative first step with probability $1/2$, for several different PDF $p(\eta)$ of the steps' lengths. The numerical data collapses exactly on the large $n$ Gaussian scaling function ${\cal G}(x)$ in Eq. \eqref{G_scal_nef}. Note that the probability that there are no records is zero as the initial position of the walk is considered as a record.}\label{Fig_G_rec}
\end{figure}

Using the result for the survival probability in Eq. \eqref{surv_res}, we compute exactly the generating function in Eq. \eqref{num_rec_GF} for the probability $R_{m,n}(q)$ that the number $N_n$ of records of the random walk up to step $n$ is exactly equal to $m$, which turns out to be also universal. In the large $n$ limit, we show that for any $0\leq q<1$, the distribution of the number of records converges to a Gaussian scaling form
\be
R_{m,n}(q)\approx\frac{1}{\sqrt{n_{\rm eff}(q)}}{\cal G}\left( \frac{m}{\sqrt{n_{\rm eff}(q)}}\right)\;,\;\;{\rm where}\;\;{\cal G}(x)=\frac{e^{-\frac{x^2}{4}}}{\sqrt{\pi}}\;.\label{G_scal_nef}
\ee
Using the Sparre-Andersen theorem, the universality of this result was already shown for an uncorrelated random walk ($q=1/2$) in \cite{MZ08}. In this expression, the parameter $n_{\rm eff}(q)$ is an effective number of steps such that $R_{m,n}(q)=R_{m,n_{\rm eff}(q)}\left(1/2\right)$, i.e. such that the probability $R_{m,n}(q)$ to have $m$ records in $n$ steps for a given value of $q$ is the same as the probability $R_{m,n_{\rm eff}(q)}\left(1/2\right)$ to have $m$ records in an effective number of steps $n_{\rm eff}(q)$ for an uncorrelated random walk ($q=1/2$). This effective number of steps reads
\be
n_{\rm eff}(q)=\frac{n}{2(1-q)}\;.\label{n_eff}
\ee
Note that the effective number is either smaller $n_{\rm eff}(q)<n$ for $q<1/2$ or bigger $n_{\rm eff}(q)>n$ for $q>1/2$ than the actual number of steps $n$ in the walk. Finally, in the limit $q\to 1$, this effective number goes to infinity. In Fig. \ref{Fig_G_rec}, we compare our analytical prediction in Eq. \eqref{G_scal_nef} with numerical simulation of the random walk. The good collapse of the data on the Gaussian scaling function indicates universality of the result as $n\to \infty$ with respect to both the PDF $p(\eta)$ and the parameter $q$. 

In the scaling regime $n\to \infty$ and $q\to 1$ with $y=n(1-q)=O(1)$, this probability converges instead to the scaling form
\begin{align}
&R_{m,n}(q)\approx\frac{1}{n}{\cal R}\left(\frac{m}{n};n(1-q)\right)\;,\label{R_scal}\\
&{\cal R}(\rho;y)=\frac{\delta(\rho)}{2}{\cal S}(y)+\frac{\delta(1-\rho)}{2}e^{-y}\\
&+\frac{e^{-y}}{2(1+\rho)}\left[y(2+\rho)\I_0(y\sqrt{1-\rho^2})+\left((y\rho-1)(1-\rho)+2y\right)\frac{\I_1(y\sqrt{1-\rho^2})}{\sqrt{1-\rho^2}}\right]\;.\nn
\end{align}
We compare in Fig. \ref{Fig_R_n} our analytical prediction for the probability of the number of records in this scaling regime to numerical data. The data shows excellent agreement with our analytical result for several distribution of the jumps' lengths $p(\eta)$.

%
%

\section{Survival Probability}\label{sec_surv}

In this Section, using the Sparre-Andersen theorem \cite{SA55}, we derive the exact expression for the survival probability $\tilde{S}^{+}_{n}(x=0;q)$, showing that it is completely universal, i.e. independent of the step distribution $p(\eta)$, for any finite $n$. The derivation below is based on the technique presented in \cite{MDMS20,MDMS20_2}, in which the survival probability of an RTP in $d$ dimensions is investigated. In the general case of starting position $x\geq 0$, one can show that $S^{+}_{n}(x;q)$ and $S^{-}_{n}(x;q)$ satisfy a set of coupled integral equations. Even if solving these equations is in general challenging, it is possible to find an exact solution for $S^{\pm}_{n}(x;q)$ in the special case of the exponential step distribution $p(\eta)=a\,e^{-a\eta}$ using a technique similar to \cite{SM12,LACTMS19}  (see Appendix \ref{app:exponential}).

\subsection{Universality of the result for $q=0$}

The first building block that we need in order to derive $S^{+}_{n}(x=0;q)$ for general $0\leq q\leq 1$ is the survival probability $S^{+}_{n}(x=0;q=0)$ for $q=0$. In this specific case, the sign of the steps is alternating at each step. Assuming that the first step is positive, each odd step will be positive while each even step is negative. Therefore if the walk survives up to step $2p$, as the $(2p+1)^{\rm th}$ step is positive it will survive for $2p+1$ steps,
\be \label{eq:odd_even}
 S_{2n+1}^{+}(x=0;q=0)=S_{2n}^{+}(x=0;q=0)\,.
 \ee  
Thus, we may restrict our analysis to the case of an even number of steps. We define the auxiliary random walk $z_p=x_{2p}$. It is easy to show that $z_p$ satisfies the relation 
\be
z_{p+1}=z_{p}+\nu_p\;,\;\;
\ee
where
\be
\nu_p=\sigma_{2p}\eta_{2p}+\sigma_{2p+1}\eta_{2p+1}=\sigma_{2p}(\eta_{2p}-\eta_{2p+1})=\eta_{2p}-\eta_{2p+1}
\ee
is a symmetric random variable of distribution given by Eq. \eqref{p_sym}. Then, since $z_p$ is a random walk with continuous and symmetric steps, one can apply the Sparre-Andersen theorem \cite{SA55}, which states that the survival probability of $z_p$ up to step $p$ is given by
\begin{equation}
S_p(0)={2p \choose p} 2^{-2p}\,.
\end{equation}
This implies that the survival probability of our original random walk with alternating-sign steps is given by
\be
S_n^{+}(x=0;q=0)=\begin{cases}
\displaystyle {2p \choose p}2^{-2p}&\;,\;\;n=2p\;,\\
\\
\displaystyle \displaystyle {2p \choose p}2^{-2p}&\;,\;\;n=2p+1\;,
\end{cases}\label{S_p_q_0}
\ee
where we have used Eq. (\ref{eq:odd_even}) in the case of odd $n$. Notably, this result in Eq. (\ref{S_p_q_0}) is valid for any $n$ and for any step distribution $p(\eta)$.

\subsection{Universality of the result in the generic case}

We are now ready to consider the general case $0\leq q \leq 1$. First of all, we define as $\tau_1\,,\cdots,\tau_m$ the successive number of steps for which the walker keeps its direction. In other words, assuming that the initial step is in the positive direction, in the first interval $k=0,\cdots,\tau_1-1$ the walker goes in the positive direction, i.e. $\sigma_k=+$, in the interval $k=\tau_1,\cdots,\tau_1+\tau_2-1$ in the negative direction, i.e. $\sigma_k=-$, and so on. Note that $1\leq m\leq n$ is a random variable, that $\tau_i\geq 1$ and that the total number of steps is fixed to be $n$
\begin{equation}
\sum_{i=1}^{m}\tau_i=n\,.
\end{equation}
The probability distribution of $\tau_i$ for $1\leq i \leq m-1$ is simply given by the geometric distribution
\begin{equation}\label{eq:f_tau}
f(\tau_i)=q^{\tau_i-1}(1-q)\,.
\end{equation}
On the other hand, the last interval $\tau_m$ has the following probability weight
\begin{equation}
q^{\tau_m-1}=\frac{1}{1-q}f(\tau_m)\,.
\end{equation}
Thus the joint probability of $\tau_1,\cdots,\tau_m$, at fixed number of steps $n$, is given by
\begin{equation}\label{joint_tau}
P(\tau_1,\ldots\,\tau_m|n)=\frac{1}{1-q}\prod_{i=1}^{m}f(\tau_i)\,\delta\left(\sum_{i=1}^{m}\tau_i-n\right)\,.
\end{equation}

Let $y_1,\cdots,y_m$ be the displacements (in absolute value) of the walker during each interval $\tau_1,\cdots,\tau_m$. Note that 
\begin{equation}
y_i=\sum_{j=0}^{\tau_i}\eta_j\,,
\end{equation}
where the $\eta_j$'s are i.i.d. variables distributed according to the continuous positive-supported PDF $p(\eta)$. Let us define the PDF of $y_i$ conditioned on the number $\tau_i$ of steps as $p_{\tau_i}(y_i)$. Even if one can in principle compute explicitly the expression of $p_{\tau}(y)$, which is the $\tau$-fold convolution of $p(\eta)$ with itself, we will see that the final result is completely independent of the specific form of $p_{\tau}(y)$, provided that it is continuous in $y$. We recall that the first displacement $\sigma_0 y_1$ is assumed to be positive, while the following displacements have alternating signs.

Using Eq. (\ref{joint_tau}) we obtain that the joint probability of $\tau_1,\cdots,\tau_m$ and $y_1,\cdots, y_m$, at fixed $n$, is 
\begin{equation}
P(\tau_1,\cdots,\tau_m,y_1,\cdots, y_m|n)=\frac{1}{1-q}\prod_{i=1}^{m}f(\tau_i)\,p_{\tau_i}(y_i)\delta\left(\sum_{i=1}^{m}\tau_i-n\right)\,.
\end{equation}
Summing over the $\tau$ variables, we get the marginal distribution of $y_1,\cdots, y_m$ at fixed total number $n$ of steps
\begin{equation}
P(y_1,\cdots, y_m|n)=\frac{1}{1-q}\prod_{i=1}^{m}\sum_{\tau_i=1}^{n}f(\tau_i)\,p_{\tau_i}(y_i)\,\delta\left(\sum_{i=1}^{m}\tau_i-n\right)\,.
\end{equation}
Taking a generating function with respect to $n$, we obtain
\begin{equation}
\sum_{n=1}^{\infty}P(y_1,\cdots, y_m|n)s^n=\frac{1}{1-q}\prod_{i=1}^{m}\sum_{\tau_i=1}^{\infty}s^{\tau_i}\,f(\tau_i)\,p_{\tau_i}(y_i)\,,
\end{equation}
which can be rewritten as
\begin{equation}\label{eq:joint_y}
\sum_{n=1}^{\infty}P(y_1,\cdots, y_m|n)s^n=\frac{1}{1-q}\left[\frac{(1-q)s}{1-qs}\right]^m\prod_{i=1}^{m}\tilde{p}_s(y_i)\,,
\end{equation}
where we have defined
\begin{equation}
\tilde{p}_s(y)=\frac{1-qs}{(1-q)s}\sum_{\tau=1}^{\infty}f(\tau)p_{\tau}(y)s^{\tau}\,.
\end{equation}
Using that $p_{\tau}(y)$ is a PDF and therefore positive and normalised to one for $y>0$, it is easy to check using the definition of $f(\tau)$ in Eq. \eqref{eq:f_tau} that the same properties are also true for $\tilde{p}_s(y)$ for any $0\leq q\leq 1$ and $0\leq s\leq 1$. Thus, $\tilde{p}_s(y)$ can be interpreted as a PDF.
It is useful to define the function
\begin{equation}
\Theta_m(y_1,\cdots, y_m)=\theta(y_1)\theta(y_1-y_2)\theta(y_1-y_2+y_3)\ldots\label{def_theta}
\end{equation}
which is one if the alternated-sign random walk with increments $y_1,\cdots, y_m$ stays above the origin up to step $m$ and it is zero otherwise.
Multiplying both sides of Eq. (\ref{eq:joint_y}) by $\Theta_m(y_1,\cdots, y_m)$, integrating over the $y$ variables and summing over $m$, we get
\begin{align}
\label{eq:joint_y2}
&\sum_{n=1}^{\infty}\sum_{m=1}^{n}\int_{0}^{\infty}dy_1\,\ldots\int_{0}^{\infty}dy_m\,\Theta_m(y_1,\cdots, y_m)P(y_1,\cdots, y_m|n)s^n\\
&=\frac{1}{1-q}\sum_{m=1}^{\infty}\int_{0}^{\infty}dy_1\,\ldots\int_{0}^{\infty}dy_m\,\Theta_m(y_1,\cdots, y_m)\left[\frac{(1-q)s}{1-qs}\right]^m\prod_{i=1}^{m}\tilde{p}_s(y_i)\,,\nonumber
\end{align}
The left-hand side of Eq. (\ref{eq:joint_y2}) is precisely the generating function of the survival probability $S^{+}_n(x=0,q)$, hence we obtain
\begin{equation}\label{eq:joint_y3}
\sum_{n=1}^{\infty}S^{+}_n(x=0,q) s^n=\frac{1}{1-q}\sum_{m=1}^{\infty}\left[\frac{(1-q)s}{1-qs}\right]^m\,Q_m\,,
\end{equation}
where we have defined 
\begin{equation}
Q_m=\int_{0}^{\infty}dy_1\,\ldots\int_{0}^{\infty}dy_m\,\Theta_m(y_1,\cdots, y_m)\prod_{i=1}^{m}\tilde{p}_s(y_i)\,.
\end{equation}

Remarkably, $Q_m$ can be interpreted as the survival probability of a random walk with  i.i.d. increments $y_1,\cdots, y_m$ drawn from the distribution $\tilde{p}_s(y)$, and with alternated increment signs (the first being positive). This corresponds precisely to the case $q=0$ considered at the beginning of this section. Thus, one has that
\begin{equation}
Q_m=S_m^{+}(x=0;q=0)
\end{equation}
where $S_m^{+}(x=0;q=0)$ is given in Eq (\ref{S_p_q_0}). Plugging this expression for $Q_m$ in Eq. (\ref{eq:joint_y3}) and using the Taylor series
\be
\sum_{k\geq 0}{{2n}\choose{n}}\left(\frac{s}{4}\right)^n=\frac{1}{\sqrt{1-s}}\;, 
\ee
we obtain, after few steps of algebra
\begin{equation}\label{eq:gen_fun}
\tilde{S}^{+}(0;s;q)=\sum_{n=0}^{\infty}S^{+}_n(x=0;q) s^n=\left(\sqrt{\frac{1-s(2q-1)}{1-s}}-q\right)\frac{1}{1-q}\,,
\end{equation}
where we have also included the term $S^{+}_0(x=0;q)=1$.
Inverting the generating function (see Appendix \ref{app:exponential} for the details), we finally obtain that for $n\geq 1$
\be \label{surv_res_2}
S_n^{+}(x=0;q)=\frac{2^{-2n}}{1-q}{2n\choose n}\pFq{2}{1}{-\frac{1}{2},-n }{\frac{1}{2}-n}{2q-1}\;,
\ee
which is indeed the result given in Eq. (\ref{surv_res}). Remarkably, this result is completely independent of the step distribution $p(\eta)$ and it is exact for any finite $n$ and for any $0\leq q \leq 1$. This survival probability $S_n^+(0,q)$ is plotted in Fig. \ref{Fig_surv_prob} for several values of $q=0,1/4,1/2,3/4$ and several distribution of steps' lengths $p(\eta)$. The numerical data for all the distributions collapses on the same master curve given by Eq. \eqref{surv_res_2}, confirming the universality of the result. The asymptotic large $n$ behaviour of the probability is most easily extracted by considering the limit $s\to 1$ of the generating function in Eq. \eqref{eq:gen_fun}, such that $\tilde S^{+}(x=0;s;q)\approx \sqrt{2/((1-q)(1-s))}$. Taking the large $n$ limit of the inverse generating function, it yields
\be
S_n^{+}(x=0;q)=\sqrt{\frac{2}{n\pi(1-q)}}+O(n^{-3/2})\;.\label{S_n_large}
\ee

\begin{figure}[h]
\centering
\includegraphics[width=0.7\textwidth]{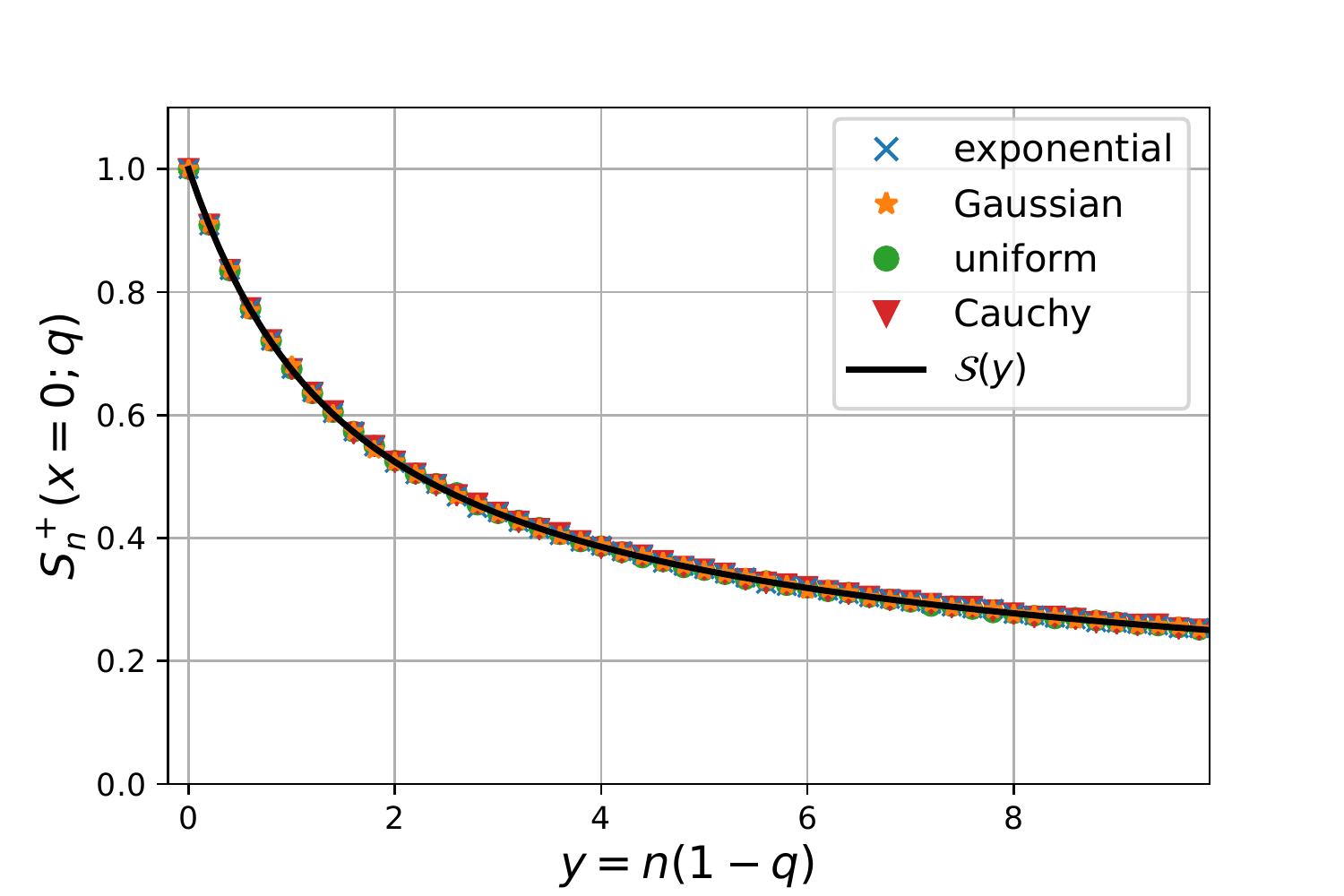}
\caption{Plot of the survival probability $S_n^{+}(x=0;q)$ as a function of the rescaled variable $y=n(1-q)$ for fixed $n=100$ and varying $q$ for an exponential (blue crosses), Gaussian (orange stars), uniform (green circles) and Cauchy (red triangles) distribution. The numerical data is obtained by simulating $N=10^5$ runs of random walks, starting with a positive first step. The numerical data collapses exactly for all the different distributions to the analytical prediction (black line) given by $S(y)$ in Eq. \eqref{scal_surv_prob}.}\label{Fig_S_y}
\end{figure}

In the limit $q\to 1$ and $n\to \infty$ with a fixed value of $y=n(1-q)$, the probability takes a universal scaling form that only depends on $y$. It can be seen by introducing in Eq. \eqref{eq:gen_fun} $q=1-y/n$ and $s=1-a/n$. Taking the large $n$ limit yields
\be
\tilde S^{+}(x=0;s;q)=\frac{n}{y}\sqrt{\frac{a+2y}{a}}+o(n)\;.
\ee
Using that $s^n=(1-a/n)^n\approx e^{-a}$, one can transform the generating function into a Laplace trnasform. The final result is then obtained by taking an inverse Laplace transform from $a\to 1$. Using the Laplace inversion formula \cite{LDMS19},
\begin{align}
{\cal L}_{a\to t}^{-1}\left(\sqrt{\frac{a+2y}{a}}e^{-x\sqrt{a(a+2y)}}\right)=&y e^{-y}\left[\I_0(y\sqrt{t^2-x^2})+\frac{t}{\sqrt{t^2-x^2}}\I_1(y\sqrt{t^2-x^2})\right]\nn\\
&+e^{-y t}\delta(t-x)\;,
\end{align}
one obtains the scaling form
\be\label{scal_surv_prob}
S_n^{+}(x=0;q)\approx {\cal S}(n(1-q))\;,\;\;{\rm with}\;\;{\cal S}(y)=e^{-y}\left(\I_0(y)+\I_1(y)\right)\;.
\ee
One can now check that in the limit $y\to \infty$, the asymptotic behaviour
\be
{\cal S}(y)\approx \sqrt{\frac{2}{\pi y}}\;,\;\;y\to \infty\;,\label{S_y_large}
\ee
matches exactly with Eq. \eqref{S_n_large} for $y=n(1-q)$. On the other hand, when $y\to 0$, one recovers the trivial limit $q=1$ where ${\cal S}(y=0)=1$. As mentioned earlier, the survival probability converges in this limit to the survival probability in continuous time for the RTP \cite{LDMS19}. A comparison between numerical data for different PDF $p(\eta)$ and the analytical prediction in Eq. \eqref{scal_surv_prob} is presented in Fig. \ref{Fig_S_y}, showing excellent agreement.



\section{Index of the maximum}\label{sec_max}

\begin{figure}[h]
\centering
\includegraphics[width=0.7\textwidth]{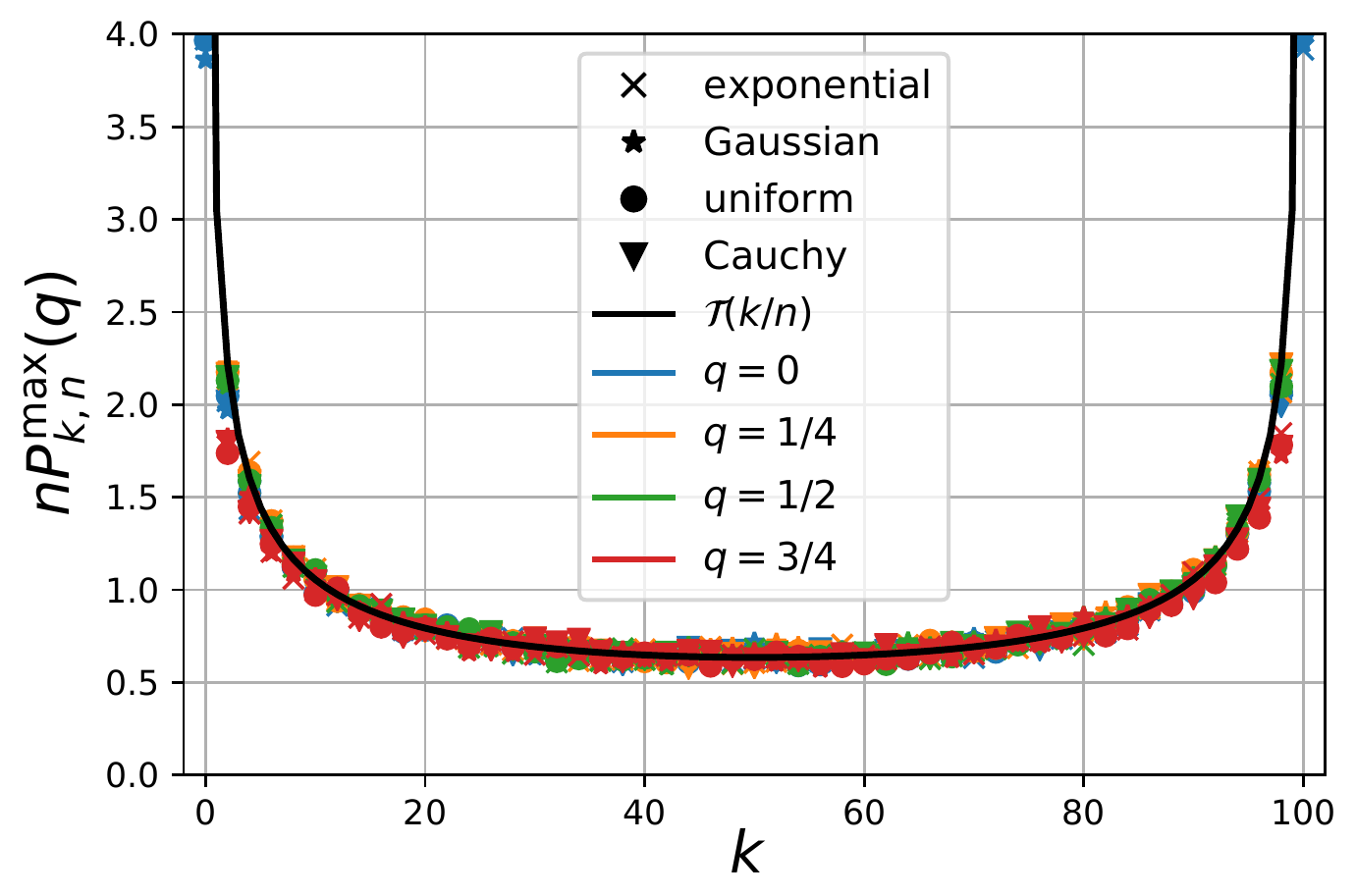}
\caption{Plot of the rescaled probability $n P_{k,n}^{\max}(q)$ of the index $n_{\max}$ of the maximum for $n=100$ and different values of $q=0,1/4,1/2,3/4$, respectively in blue, orange, green and red as a function of $k=0,\cdots,100$. The numerical data is obtained by simulating $N=10^5$ runs of random walks starting with a positive or negative first step with probability $1/2$, for several PDF $p(\eta)$ of the steps' lengths. The numerical data collapse on the limit scaling function ${\cal T}(\tau)$ in Eq. \eqref{arcsine}.}\label{Fig_T_max_large}
\end{figure}

\begin{figure}[h]
\centering
\includegraphics[width=0.7\textwidth]{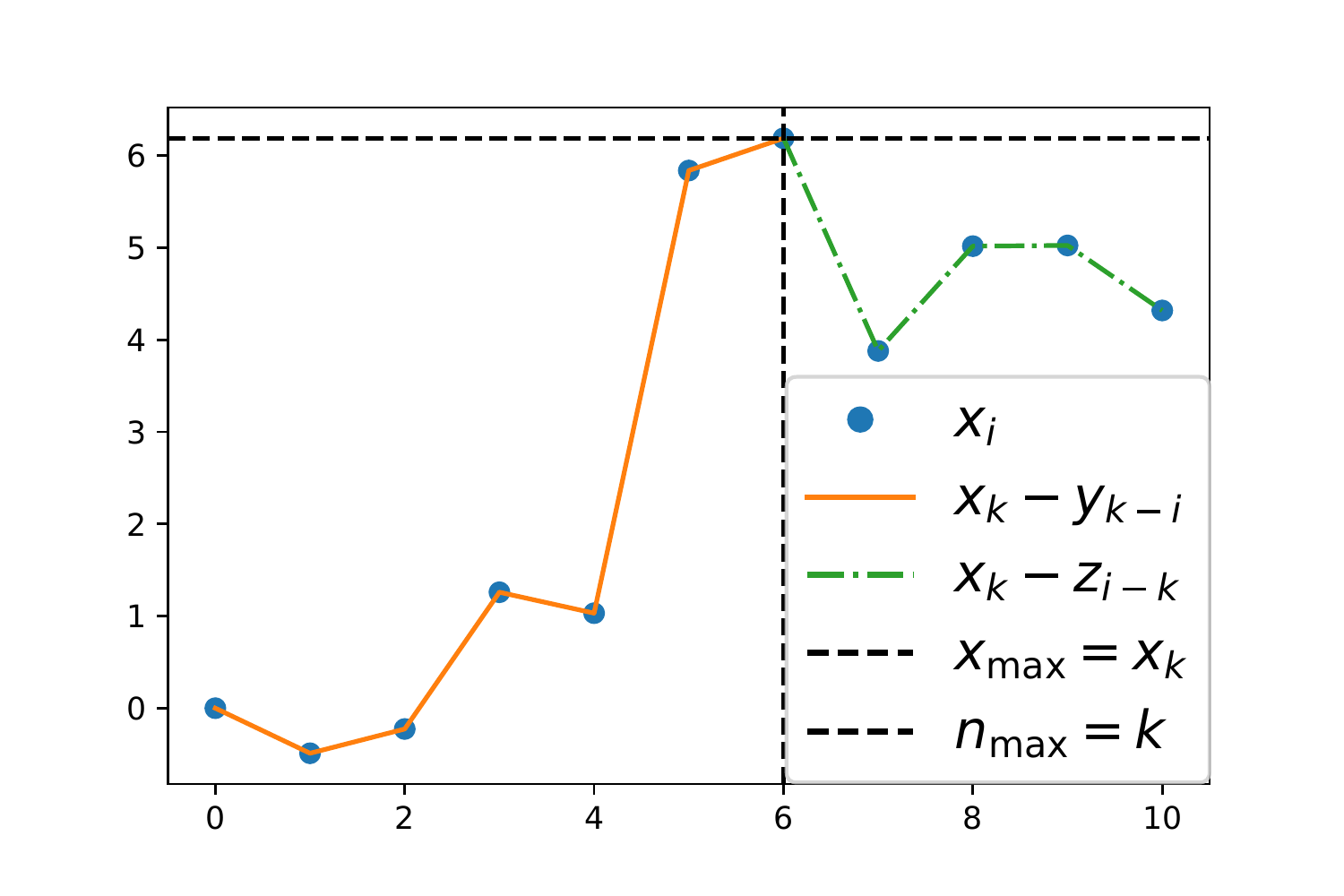}
\caption{Sketch of the decomposition of a random walk $x_i$ for $i=0,\cdots,n$ into two smaller random walk $y_j$ for $j=0,\cdots,k$ and $z_l$ for $l=0,\cdots,n-k$, where $k=n_{\max}$ is the index at which $x_k=x_{\max}=\max_i x_i$. As the maximum is reached at step $k$, the random walks $y_j$ and $z_l$ respectively survive for $k$ and $n-k$ steps.}\label{Fig_decomp_sketch}
\end{figure}

The formula \eqref{P_max_n} is exact for any finite $n$ and gives the probability $P_{k,n}^{\max}$ of the index $n_{\max}$ of the maximum from the survival probability $S_{k}^+(x=0;q)$. To understand this formula, we first suppose that the index of the maximum is $0<n_{\max}=k<n$. We separate our initial walk $x_i$ for  $i=0,\cdots, n$ into two smaller walks defined as (see also Fig. \ref{Fig_decomp_sketch})
\be
y_j=x_k-x_{k-j}\;,\;\;i=0,\cdots,k\;,\;\;z_l=x_k-x_{k+l}\;,\;\;l=0,\cdots,n-k\;.
\ee
As $x_k$ is the maximum of the random walk, one needs to have $y_j\geq 0$ for all $j=0,\cdots,k$ and $z_l\geq 0$ for all $l=0,\cdots,n-k$. The probability of the latter event is simply given by the survival probability $S_{n-k}^+(x=0;q)$. As for the probability of the former event, i.e. $y_j\geq 0$ for all $j=0,\cdots,k$, one needs to consider the probability of a reverse trajectory of the random walk defined in Eqs. \eqref{RW} and \eqref{sigma_def}. First, let us note that for a random walk with positive or negative first step with equal probability $\Prob(\sigma_0=+)=\Prob(\sigma_0=-)=1/2$, any trajectory in the state space of the $\sigma_k$'s has the same probability weight as its reverse trajectory
\be
P(\sigma_0\to\sigma_1\to\sigma_2\to\cdots\sigma_{n-1}\to\sigma_n)=P(\sigma_n\to\sigma_{n-1}\to\cdots\sigma_2\to\sigma_1\to \sigma_0).
\ee
To show this, we use the fact that the process is Markovian
\begin{align}
&P(\sigma_0\to\sigma_1\to\sigma_2\to\cdots\sigma_{n-1}\to\sigma_n)\nn\\ 
&=\Prob(\sigma_0)\Prob(\sigma_1|\sigma_0)\Prob(\sigma_2|\sigma_1)\cdots \Prob(\sigma_n|\sigma_{n-1})\;.
\end{align}
As $\sigma_0$ is equal to $\pm$ with equal probability $1/2$, one can show that
\be
\Prob(\sigma_1=+)=q \Prob(\sigma_0=+)+(1-q)\Prob(\sigma_0=-)=\Prob(\sigma_0=+)=\frac{1}{2}\;,
\ee
and by recursion that $\Prob(\sigma_k=+)=\Prob(\sigma_k=-)=1/2$ for all $0\leq k\leq n$. Next, we use Bayes' theorem to obtain that
\be
\Prob(\sigma_{k}|\sigma_{k+1})=\Prob(\sigma_{k+1}|\sigma_{k})\frac{\Prob(\sigma_k)}{\Prob(\sigma_{k+1})}=\Prob(\sigma_{k+1}|\sigma_{k})\;.
\ee
Using this result, it is trivial to check that
\begin{align}
&P(\sigma_0\to\sigma_1\to\sigma_2\to\cdots\sigma_{n-1}\to\sigma_n)\\
&=\Prob(\sigma_0)\Prob(\sigma_1|\sigma_0)\Prob(\sigma_2|\sigma_1)\cdots \Prob(\sigma_n|\sigma_{n-1})\nn\\
&=\Prob(\sigma_n)\Prob(\sigma_n|\sigma_{n-1})\Prob(\sigma_{n-2}|\sigma_{n-1})\cdots \Prob(\sigma_0|\sigma_{1})\nn\\
&=P(\sigma_n\to\sigma_{n-1}\to\cdots\sigma_2\to\sigma_1\to \sigma_0)\;.\nn
\end{align}
Note that, on the other hand, the lengths $\eta_i$'s are i.i.d. random variables and the weight of any random walk is invariant by permutation of the lengths of its increments. We can now use that the probability that the reverse random walk $y_j$ survives, i.e. $y_j\geq 0$ for all $j=0,\cdots,k$, with a positive or negative last step with equal probability $1/2$ is the same as the probability that a random walk starting with a positive or negative first step with equal probability $1/2$ survives and is therefore simply given by $(1/2)S_k^+(x=0;q)$. In order for $x_k$ to be the maximum, one needs both that $x_k>x_{k-1}$ and $x_{k}>x_{k+1}$, such that $\sigma_{k-1}=\sgn(x_k-x_{k-1})=+$ while $\sigma_{k+1}=\sgn(x_{k+1}-x_{k})=-$, which occurs with probability $1-q$. Apart from this factor, the two random walks $y_j$'s and $z_l$'s are totally independent. The probability $P_{k,n}^{\max}$ is thus the product of the survival probabilities for each walk and of $1-q$. 

If on the other hand, the index of the maximum is $k=0$ or $k=n$, the first step must be negative (resp. the last step is positive) which happens with probability $1/2$. All steps of the random walk must then remain below $x_0$ (resp. $x_n$) for $n$ step, which occurs with probability $S_n^+(x=0;q)$.

In order to check that the probability distribution $P_{k,n}^{\max}$ is correctly normalized to one, we consider the double generating function of $P_{k,n}^{\max}$ with respect to $n\geq 1$ and $0\leq k \leq n$. Using Eqs. (\ref{P_max_n}) and (\ref{eq:gen_fun}) we get, after few steps of algebra,
\be
\sum_{n=1}^{\infty}\sum_{k=0}^{n}P_{k,n}^{\max}z^k\,s^n=\frac{1}{2(1-q)}\left(\sqrt{\frac{1-s(2q-1)}{1-s}}-1\right)\left(\sqrt{\frac{1-s z(2q-1)}{1-s z}}-1\right)\,.
\ee
Setting $z=1$ on both sides, we obtain
\be
\sum_{n=1}^{\infty}\sum_{k=0}^{n}P_{k,n}^{\max}\,s^n=\frac{s}{1-s}\,.
\ee
And finally, inverting the generating function with respect to $n$, we obtain that for any $n\geq 1$,
\begin{equation}
\sum_{k=0}^{n}P_{k,n}^{\max}=1\,.
\end{equation}

In the large $n$ limit with $0<\tau=k/n<1$ and $0\leq q<1$, one can replace the survival probabilities in Eq. \eqref{P_max_n} by their asymptotic behaviour in Eq. \eqref{S_n_as}. One then realises that the $q$ dependency vanishes and one is left with the arcsine scaling form in Eq. \eqref{arcsine}. The universality of this result in the large $n$ limit with respect to both the distribution $p(\eta)$ and the value of $q$ is verified numerically in Fig. \ref{Fig_T_max_large}.   

\begin{figure}[h]
\centering
\includegraphics[width=0.7\textwidth]{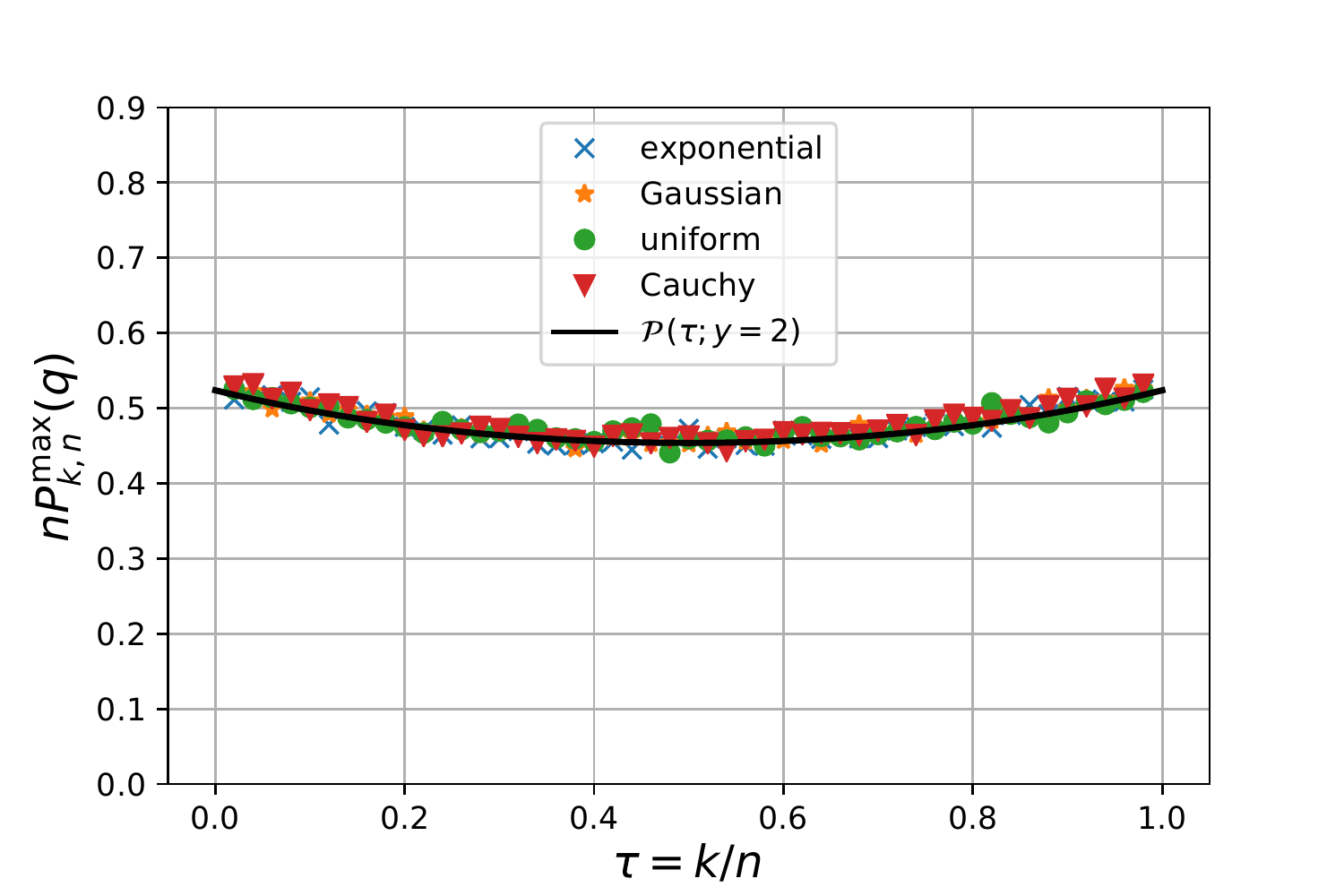}
\caption{Plot of the rescaled probability $n P^{\max}_{k,n}(q)$ of the index of the maximum as a function of the rescaled variable $\tau=k/n$ for fixed $n=100$ and $y=n(1-q)=2$ for an exponential (blue crosses), Gaussian (orange stars), uniform (green circles) and Cauchy (red triangles) distribution. The numerical data is obtained by simulating $N=10^6$ runs of random walks, starting with a positive or negative first step with probability $1/2$. The numerical data collapses exactly for all the different distributions to the analytical prediction (black) given by ${\cal P}(\tau;y)$ in Eq. \eqref{P_max_RTP}. Note that the delta peaks for $\tau=0$ and $\tau=1$ in the expression in Eq. \eqref{P_max_RTP} are not shown for convenience but are indeed obtained in the numerical data.}\label{Fig_P_scal}
\end{figure}

In the scaling limit $n\to \infty$ and $q\to 1$ with $y=n(1-q)=O(1)$, one can just replace in Eq. \eqref{P_max_n} the survival probabilities by their asymptotic behaviour in Eq. \eqref{scal_surv_prob}. This yields the scaling form in Eq. \eqref{P_max_RTP}. Note that using the asymptotic behaviour for large $y$ in Eq. \eqref{S_y_large}, one recovers form Eq. \eqref{P_max_RTP} the arcsine scaling function in this limit. In Fig. \ref{Fig_P_scal}, we check the universality of this result by comparing our analytical result in this scaling limit and the result from numerical simulations for different distributions of the jumps' lengths $p(\eta)$.

\section{Record statistics}\label{sec_rec}

We now consider the record statistics for this random walk of $n$ steps. Similarly to the standard definition \cite{SM14}, the initial position is counted as a record. The random walk reaches a record at step $k\geq 1$ if 
\be
x_{k}=\max_{0\leq i\leq k}x_i.
\ee
The age of a record is defined as the number of steps for which this record is standing \cite{SM14} (see Fig. \ref{Fig_sketch_rec}). While all other records are standing up to the step where a new record is reached, the last records stands up to the final step $n$ is reached. 

\subsection{Joint probability of the ages}

We start by deriving the joint probability of the ages $l_1,\cdots,l_m$ of the records, given that there are $N_n=m$ records in $n$ steps. We first note that if a record is reached at step $l$, one must have $x_l>x_{l-1}$ such that $\sigma_{l-1}=\sgn(x_l-x_{l-1})=+$. Using this feature, one can consider independently the different parts of the random walk in between each record by defining $m$ smaller random walks
\be
y^k_{i}=x_{n_k}-x_{n_k+i}\;,\;\;n_1=0\;\;,\;\;n_k=\sum_{i=1}^{k-1} l_i\;,\;\;k=1\;,\;\cdots\;,m\;.
\ee
The $k$-th record (with $1<k<m$) is broken after $l$ steps if $y_i^{k}<0$ for $i=0,\cdots,l-1$ and $y_l^k>0$. This event occurs with the first passage probability
\be
F_l(0;q)=S_{l-1}(0;q)-S_l(0;q)\;,\;\;l\geq 1\;,\label{F_l}
\ee
where the survival probability $S_l(0;q)$ is defined as
\be
S_l(0;q)=(1-q)S_l^+(0;q)+q S_l^-(0;q)\;.
\ee
As the final step before reaching a record is always positive, the first step $y^k_1$ of the $k^{\rm th}$ walk for any $1\leq k\leq m$ is negative with probability $q$ and positive with probability $1-q$ (the steps of $y^k$ are of opposite sign as those of the original walk $x_i$).
For the first record instead, the first step is positive or negative with probability $1/2$ and the probability that the record is broken after $l_1$ steps is given by the symmetric first passage probability
\begin{align}
F_{l_1}^{\rm sym}(0;q)&=S_{l_1-1}^{\rm sym}(0;q)-S_{l_1}^{\rm sym}(0;q)\;,\;\;l_1\geq 1\;,\\
S_{l_1}^{\rm sym}(0;q)&=\frac{1}{2}\left[S_{l_1}^+(0;q)+S_{l_1}^-(0;q)\right]\;.
\end{align}
The last record is not broken at the final step $n$ such that the probability for the last age $l_m$ is just the survival probability $S_{l_m}(0;q)$. Note that if one has only one record in the random walk, it means that the random walk never reaches a position bigger than $x_0=x=0$. This event occurs with probability $S_{n}^{\rm sym}(x=0;q)$. The joint probability of the ages of the records, starting with a positive or negative first step with probability $1/2$ therefore reads for this process
\be\label{P_j_q_RW}
P_{m,n}(l_1,\cdots,l_m;q)=\begin{cases}
\displaystyle S_{n}^{\rm sym}(0;q)&\;,\;\;m=1\;,\\
&\\
\displaystyle F_{l_1}^{\rm sym}(0;q)\prod_{k=1}^{m-2}F_{l_k}(0;q)S_{l_m}(0;q)\delta_{n,\sum_{i=1}^{m}l_i}&\;,\;\;m\geq 2\;.
\end{cases}
\ee
Notably, Eq. (\ref{P_j_q_RW}) shows that the full record statistics is universal for this class of random walks. Note that this expression in Eq. (\ref{P_j_q_RW}) is very similar to the expression for uncorrelated random walks \cite{MZ08}. It is exactly recovered in the particular case where $q=1/2$ for which $F_{l}(0;1/2)=F_{l}^{\rm sym}(0;1/2)$. For $q=1$, this expression becomes trivial. The first passage probabilities are given by $F_{l}(0;1)=2F_{l}^{\rm sym}(0;1)=\delta_{l,1}$, while the survival probabilities are $S_{l}(0;1)=2S_l^{\rm sym}(0;1)=1$. If the first step is positive (resp. negative), which occurs with probability $1/2$, all the steps are positive (resp. negative) and there are $n$ records whose ages are $l=1$ (resp. a single record whose age is $l=n$).

\subsection{Number of records}

Using the joint probability of the ages of the record in Eq. \eqref{P_j_q_RW}, one can obtain an exact expression for the generating function of the probability $R_{m,n}(q)$ that there are $N_n=m$ records in $n$ steps.  Multiplying Eq. \eqref{P_j_q_RW} by $s^n$, summing over $n$ and all the possible values $l_i>1$ for $i=1,\cdots,n$ yields
\begin{align}
&\tilde R_m(s;q)=\sum_{n\geq 0}s^n R_{m,n}(q)
&=\begin{cases}
\displaystyle \tilde S^{\rm sym}(0;s;q)&\;,\;\;m=1\;,\\
&\\
\displaystyle \tilde F^{\rm sym}(0;s;q)\tilde F(0;s;q)^{m-2}\tilde S(0;s;q)&\;,\;\;m\geq 2\;,
\end{cases}\label{num_rec_GF}
\end{align}
where $\tilde S(0;s;q),\,\tilde S^{\rm sym}(0;s;q),\,\tilde F(0;s;q),\,$and $\tilde F^{\rm sym}(0;s;q)$ indicate the generating functions with respect to $n$ of $ S_n(0;q),\, S_n^{\rm sym}(0;q),\, F_n(0;q),\,$and $ F_n^{\rm sym}(0;q)$, respectively.
Note that using Eq. \eqref{F_l}, there is a simple identity between the generating function of the first passage and survival probabilities
\be
\tilde F^{\pm}(0;s;q)=1-(1-s)\tilde S^{\pm}(0;s;q)\;.
\ee

\subsubsection{Number of records in the large $n$ limit}\hfill\\

In the large $n$ limit and for $0\leq q<1$, one can obtain the probability of the number of records from Eq. \eqref{num_rec_GF}. We introduce the rescaled variable $s=1-a/n$ and suppose that in large $n$ limit, the number of records scales as $m=O(\sqrt{n})$. Taking the large $n$ limit in Eq. \eqref{num_rec_GF}, we obtain
\be
\tilde R_m\left(s=1-\frac{a}{n};q\right)=\sqrt{\frac{2n(1-q)}{a}}e^{-\frac{m}{\sqrt{n}}\sqrt{2a(1-q)}}+O(1)\;.
\ee
In this large $n$ limit, one can replace the generating function inversion by a Laplace transform inversion from $a\to 1$. It can be taken explicitly using the formula
\be
{\cal L}_{a\to t}^{-1}\left(\frac{e^{-x\sqrt{a}}}{\sqrt{a}}\right)=\frac{e^{-\frac{x^2}{4t}}}{\sqrt{\pi t}}\;.
\ee
The distribution of the number of records in the large $n$ limit takes the following scaling form
\be\label{G_scal_form}
R_{m,n}(q)\approx \sqrt{\frac{2(1-q)}{n}}{\cal G}\left( \sqrt{\frac{2(1-q)}{n}}m\right)\;,\;\;{\rm where}\;\;{\cal G}(x)=\frac{e^{-\frac{x^2}{4}}}{\sqrt{\pi}}\;.
\ee
Note that the scaling form is universal and the correlation parameter $q$ only appears as a rescaling parameter. In fact, as previously noticed, using the result for uncorrelated random walks (corresponding to the case $q=1/2$), in the large $n$ limit the number of records has the same distribution as an uncorrelated random walk with an effective number of steps $
n_{\rm eff}(q)=n/(2(1-q))$. We have checked numerically in Fig. \ref{Fig_G_rec} that the scaling form in Eq. \eqref{G_scal_form} is indeed universal with respect to both the PDF of the jumps' lengths $p(\eta)$ and the parameter $0\leq q<1$.

 \subsubsection{Scaling function in the limit $q\to 1$}\hfill\\

\begin{figure}[h]
\centering
\includegraphics[width=0.7\textwidth]{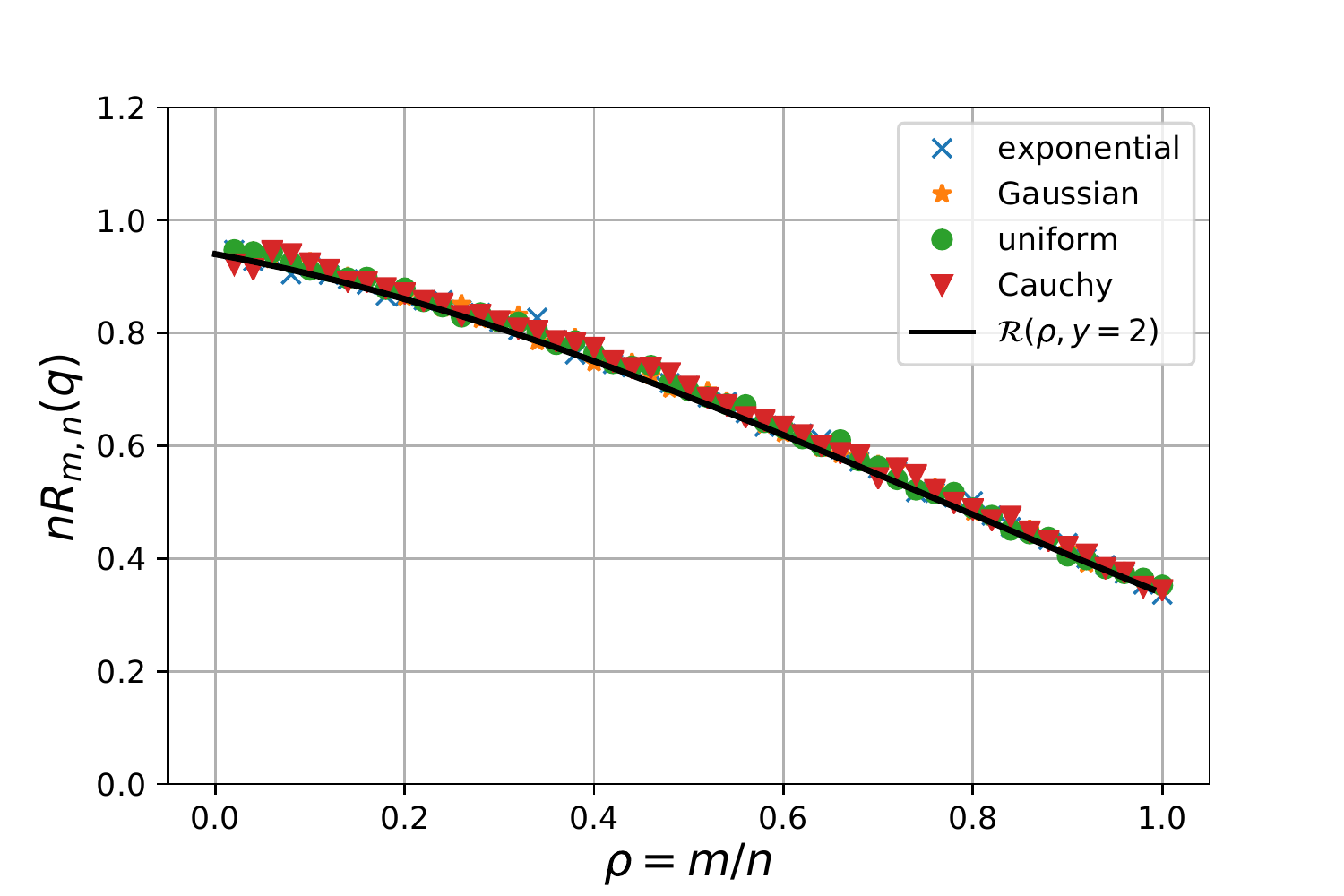}
\caption{Plot of the rescaled probability $n R_{m,n}(q)$ as a function of the rescaled variable $\rho=m/n$ for fixed $n=100$ and $y=n(1-q)=2$ for an exponential (blue crosses), Gaussian (orange stars), uniform (green circles) and Cauchy (red triangles) distribution. The numerical data is obtained by simulating $N=10^5$ runs of random walks, starting with a positive/negative first step with probability $1/2$. The numerical data collapses exactly for all the different distributions to the analytical prediction (black) given by ${\cal R}(\rho;y)$ in Eq. \eqref{R_scal_form}. Note that the delta peaks for $\rho=0$ and $\rho=1$ in the expression in Eq. \eqref{R_scal_form} are not shown for convenience but are indeed obtained in the numerical data.}\label{Fig_R_n}
\end{figure}

We consider now the scaling regime $n\to \infty$ and $q\to 1$ with $y=n(1-q)=O(1)$. In this scaling regime, we anticipate that the number of records $m=O(n)$ instead of $O(\sqrt{n})$ as obtained for $q<1$. We introduce the rescaled variables $s=1-a/n$, and take the large $n$ limit in Eq. \eqref{num_rec_GF} with fixed $a,y,m/n=O(1)$. It yields
\begin{align}
&\tilde R_m\left(s=1-\frac{a}{n};q=1-\frac{y}{n}\right)\approx \frac{n}{2y}\left(\sqrt{\frac{a+2y}{a}}-1\right)\\
&+\frac{n}{2y}\sqrt{\frac{a+2y}{a}}\left[2y+a-\sqrt{a(a+2y)}\right]e^{-\frac{m}{n}\sqrt{a(a+2y)}}+O(1)\;.
\end{align}
In this large $n$ limit, the generating functions can be inverted by taking the inverse Laplace transform from $a\to 1$. Using the inverse Laplace formulae \cite{LDMS19}
\begin{align}
{\cal L}_{a\to t}^{-1}\left(e^{-x\sqrt{a(a+2y)}}\right)=&\frac{yx e^{-y t}}{\sqrt{t^2-x^2}}\I_1(y\sqrt{t^2-x^2})+e^{-y t}\delta(t-x)\;,\\
{\cal L}_{a\to t}^{-1}\left(\frac{\sqrt{a(a+2y)}}{2y}e^{-x\sqrt{a(a+2y)}}\right)=&-\frac{1}{2y}\partial_x {\cal L}_{a\to t}^{-1}\left(e^{-x\sqrt{a(a+2y)}}\right)\;,\\
{\cal L}_{a\to t}^{-1}\left(\frac{a}{2y}e^{-x\sqrt{a(a+2y)}}\right)=&\frac{1}{2y}\partial_t {\cal L}_{a\to t}^{-1}\left(e^{-x\sqrt{a(a+2y)}}\right)\;,
\end{align}
we obtain the final scaling form of the probability of the number of records in this limit
\begin{align}
&R_{m,n}(q)\approx\frac{1}{n}{\cal P}\left(\frac{m}{n};n(1-q)\right)\;,\label{R_scal_form}\\
&{\cal R}(\rho;y)=\frac{\delta(\rho)}{2}e^{-y}\left[\I_0(y)+\I_1(y)\right]+\frac{e^{-y}}{2}\delta(1-\rho)\\
&+\frac{e^{-y}}{2(1+\rho)}\left[y(2+\rho)\I_0(y\sqrt{1-\rho^2})+\left((y\rho-1)(1-\rho)+2y\right)\frac{\I_1(y\sqrt{1-\rho^2})}{\sqrt{1-\rho^2}}\right]\;.
\end{align}
In the limit $y\to \infty$, $\rho\to 0$ with $\sqrt{y}\rho=O(1)$, and using the asymptotic behaviour \cite{DLMF_I_as}
\be
e^{-y}\I_{\nu}(y\sqrt{1-\rho^2})\approx \frac{e^{-\frac{y \rho^2}{2}}}{\sqrt{2\pi y}}\;,\;\;\nu=O(1)\;,
\ee
one obtains that the scaling function ${\cal R}(\rho;y)$ asymptotically behaves as
\be
{\cal R}(\rho;y)\approx \sqrt{2y}\,{\cal G}\left(\sqrt{2y}\rho\right)\;,\;\;y\to \infty\;,\;\;\rho\to 0\;\;,\;\sqrt{y}\rho=O(1)\;.
\ee 
Inserting this asymptotic behaviour in Eq. \eqref{R_scal_form}, we recover the scaling form for $q<1$ and large $n$ in Eq. \eqref{G_scal_form}. In Fig. \ref{Fig_R_n}, we have checked numerically that the scaling form in Eq. \eqref{R_scal_form} is universal with respect to the PDF $p(\eta)$ for a fixed value of $n$ and $y=n(1-q)$.

%
%
%
\section{Conclusion}\label{sec_conc}

In this article, we have introduced a model of one-dimensional random walk with correlated steps such that the sign of consecutive steps is the same with probability $q$, with $0\leq q\leq 1$ a parameter controlling the persistence of the random walk. We have computed analytically the survival probability starting from the origin for any PDF $p(\eta)$ of the steps' lengths, showing that it is universal. As a first application of this result, we have derived the probability distribution for the step $n_{\max}$ at which the walk reaches its maximum in Eq. \eqref{P_max_n}. Using again our result for the survival probability as a building block, we have derived the joint probability for the ages of the records in Eq. \eqref{P_j_q_RW}. Finally, using this result we have derived the distribution for the number $N_n$ of records in the large $n$ limit. We have shown that as $n\to \infty$, there are two distinct scaling regimes for the properties of this random walk and that these regimes match smoothly with one another. In the regime where $0\leq q<1$ is fixed as $n\to \infty$, the record statistics are identical as that of an uncorrelated random walk with an effective number of steps $n_{\rm eff}(q)=n/(2(1-q))$. In the regime where $n\to \infty$ and $q\to 1$ with $y=n(1-q)$, the properties of the random walk are the same as for the run-and-tumble particle. 

For a continuous and symmetric random walk, the distribution of: (i) the step $n_{\max}$ at which the random walk reaches its maximum and (ii) the number of steps that the random walk has spent above the origin are both given by the same law $P_{k,n}=2^{-2n}{{2k}\choose {k}}{{2(n-k)}\choose {n-k}}$ and are both universal with respect to the step distribution. We have argued here that the law for $n_{\max}$ remains universal for our model. It could be interesting to check whether the law for the number of steps that the random walk has spent above the origin is also universal and given by the same law as $n_{\max}$. Preliminary numerical simulation indicate that while the probability for the random variable (ii) is given by $P_{k,n}^{\max}$ in the particular case of the exponential distribution, this variable is not universal with respect to $p(\eta)$.

We could also readily extend the results obtained here to study the extreme value, order and gap statistics of the correlated random walk as it was obtained for standard symmetric continuous random walk in \cite{SM12,LACTMS19}. This is left for future investigations.

\ack We thank S.N. Majumdar, D. Mukamel, and G. Schehr for their helpful comments and for pointing out useful references.

\newpage

\appendix

\section{Another method to derive the survival probability}\label{app:exponential}
In order to compute the survival probability, we introduce the pair of backward equation for the probability $S_n^\sigma(x;q)$ of surviving for $n$ steps starting from a position $x$ with a positive $\sigma=+$ (resp. negative $\sigma=-$) first step. It reads for $x\geq 0$,
\begin{align}
S_{n+1}^{+}(x;q)&=\int_x^{\infty}dy\, p(y-x)\left[q S_n^+(y;q)+(1-q)S_n^-(y;q)\right]\;,\label{S_n_p}\\
S_{n+1}^{-}(x;q)&=\int_0^{x}dy\, p(x-y)\left[(1-q) S_n^+(y;q)+q S_n^-(y;q)\right]\;.\label{S_n_m}
\end{align}
In this expression, starting from a positive position $x$, the walk survives on the first step if it goes to a position $y>0$. The step's length $\eta=|x-y|$ is a random variable of PDF $p(\eta)$. The sign of the second step will be $\sigma$, i.e. the same as the first step, with probability $q$ and $-\sigma$, i.e. opposite to the first step, with probability $1-q$.
To solve this equation, we first introduce the generating function
\be
\tilde S^{\sigma}(x;s;q)=\sum_{n=0}^{\infty}S_n^{\sigma}(x;q) s^n\;. 
\ee
Multiplying Eqs. \eqref{S_n_p} and \eqref{S_n_m} by $s^{n+1}$ and summing over $n>0$, we obtain a set of two closed integral equations for the generating functions
\begin{align}
\tilde S^{+}(x;s;q)&=1+s\int_x^{\infty}dy\, p(y-x)\left[q \tilde S^{+}(y;s;q)+(1-q)\tilde S^{-}(y;s;q)\right]\;,\label{GF_S_+}\\
\tilde S^{-}(x;s;q)&=1+s\int_0^{x}dy\, p(x-y)\left[(1-q) \tilde S^{+}(y;s;q)+q \tilde S^{-}(y;s;q)\right]\label{GF_S_-}\;.
\end{align}
In this expression, we used the trivial condition $S_{0}^{\sigma}(x)=\Prob\left[x_0=x\geq 0\right]=\Theta(x)$, where $\Theta(x)$ is the Heaviside step-function.  

\subsection{Exponential distribution}

For the exponential distribution, one can solve exactly for the full distribution $S_{n}^{\sigma}(x;q)$. To show this, we first derive Eqs. \eqref{GF_S_+} and \eqref{GF_S_-} with respect to $x$,
\begin{align}
\partial_x\tilde S^{+}(x;s;q)=&-s p(0)\left[q \tilde S^{+}(x;s;q)+(1-q)\tilde S^{-}(x;s;q)\right]\nn\\
&+s\int_x^{\infty}dy\, \partial_x p(y-x)\left[q \tilde S^{+}(y;s;q)+(1-q)\tilde S^{-}(y;s;q)\right]\;,\label{d_S_p}\\
\partial_x\tilde S^{-}(x;s;q)=&s p(0)\left[(1-q) \tilde S^{+}(x;s;q)+q\tilde S^{-}(x;s;q)\right]\nn\\
&+s\int_0^{x}dy\, \partial_x p(x-y)\left[(1-q) \tilde S^{+}(y;s;q)+q \tilde S^{-}(y;s;q)\right]\;.\label{d_S_m}
\end{align}
The exponential distribution satisfies the simple relation
\be
p(\eta)=a e^{-a \eta}\;,\;\;\partial_\eta p(\eta)=-a p(\eta)\;.
\ee 
Inserting in Eqs. \eqref{d_S_p} and \eqref{d_S_m}, it yields the following set of coupled first order differential equations
\begin{align}
\partial_x\tilde S^{+}(x;s;q)=&-a s\left[q \tilde S^{+}(x;s;q)+(1-q)\tilde S^{-}(x;s;q)\right]+a\tilde S^{+}(x;s;q)-a\;,\\
\partial_x\tilde S^{-}(x;s;q)=& a s \left[(1-q) \tilde S^{+}(x;s;q)+q\tilde S^{-}(x;s;q)\right]-a\tilde S^{+}(x;s;q)+a\;.
\end{align}
The solution of these equations that does not diverge as $x\to \infty$ reads
\be
\tilde S^{\sigma}(x;s;q)=\frac{1}{1-s}+A^{\sigma}(s;q)e^{-a x \sqrt{(1-s)(1-(2q-1)s)}}\;.\label{ansatz}
\ee
In order to find the coefficients $A^{\pm}(s;q)$, we reintroduce this solution in the integral equations \eqref{GF_S_+} and \eqref{GF_S_-} . This yields
\begin{align}
A^{+}(s;q)e^{-a x \sqrt{(1-s)(1-(2q-1)s)}}=&s\frac{q A^{+}(s;q)+(1-q)A^{-}(s;q)}{1+\sqrt{(1-s)(1-(2q-1)s)}}e^{-a x \sqrt{(1-s)(1-(2q-1)s)}}\\
A^{-}(s;q)e^{-a x \sqrt{(1-s)(1-(2q-1)s)}}=&s\frac{q A^{-}(s;q)+(1-q)A^{+}(s;q)}{1-\sqrt{(1-s)(1-(2q-1)s)}}e^{-a x \sqrt{(1-s)(1-(2q-1)s)}}\\
&-s\left[\frac{1}{1-s}+\frac{q A^{-}(s;q)+(1-q)A^{+}(s;q)}{1-\sqrt{(1-s)(1-(2q-1)s)}}\right]e^{-a x}\;.
\end{align}
Identifying the constant pre-exponential coefficients of $e^{-a x}$ and $e^{-a x \sqrt{(1-s)(1-(2q-1)s)}}$ on each side of the equation, the system of algebraic equations can be solved, yielding
\be
A^{-}(s;q)=-\frac{s}{1-s}\;,\;\;A^{+}(s;q)=\frac{1}{1-q}\left[-\frac{1-q s}{(1-s)}+\sqrt{\frac{1-(2q-1)s}{1-s}}\right]\;.\label{A_s_q}
\ee
Note that the coefficient $A^{-}(s;q)$ does not depend on $q$. Inserting this result in Eq. \eqref{ansatz}, one obtains that $\tilde S^{-}(x=0;s;q)=1$ from which one can extract easily the trivial relation $S_n^{-}(x=0;q)=\delta_{n,0}$. One can also compute the generating function of the survival probability for $x=0$ starting in state $+$, which reads
\be
\tilde S^{+}(x=0;s;q)=\frac{1}{1-s}+A^{+}(s;q)=\frac{1}{1-q}\left[-q+\sqrt{\frac{1-(2q-1)s}{1-s}}\right]\,,\label{GF_S_p}
\ee
which is in agreement with the result in Eq. (\ref{eq:gen_fun}).
Using the Taylor series
\be
(1-x)^a=\sum_{k=0}^{\infty}{{a}\choose {k}}(-x)^k\;,
\ee
one can extract the probability $S_n^{+}(x=0;s;q)$ from this expression
\be
S^{+}(x=0;s;q)=1+\frac{(-1)^n}{1-q}\sum_{n=1}^{\infty} s^n \sum_{k=0}^n {-\frac{1}{2}\choose n-k}{\frac{1}{2}\choose k}(2q-1)^k\;.\label{S_int}
\ee
To obtain the final result in Eq. \eqref{surv_res}, we use the identities \cite{DLMF1,DLMF2}
\begin{align}
&{a\choose k}=(-1)^k {k-a-1\choose k}\;,\;\;{a\choose k}= \frac{(a+1-k)_k}{k!}=\frac{(-1)^k}{k!}(-a)_k\;,\\
&(a)_{n+k}=(a)_n(a+n)_k\;,\;\;(a)_{-n}=\frac{1}{(a-n)_n}\;,
\end{align}
where $(a)_k=\Gamma(a+k)/\Gamma(a)$ is the rising factorial (or Pochhammer symbol). We can then use these identities to rewrite the sum in Eq. \eqref{S_int} as
\begin{align}
&\sum_{k=0}^n  {-\frac{1}{2}\choose n-k}{\frac{1}{2}\choose k}(x)^k= \frac{1}{n!}\sum_{k=0}^n {n\choose k}\left(-\frac{1}{2}\right)_{k}\left(\frac{1}{2}-n+k\right)_{n-k}(-x)^k\\
&=\frac{\left(\frac{1}{2}-n\right)_{n}}{n!}\sum_{k=0}^n {n\choose k}\left(-\frac{1}{2}\right)_{k}\left(\frac{1}{2}-n+k\right)_{-k}(-x)^k\\
&={-\frac{1}{2}\choose n}\sum_{k=0}^n {n\choose k}\frac{\left(-\frac{1}{2}\right)_{k}}{\left(\frac{1}{2}-n\right)_{k}}(-x)^k={-\frac{1}{2}\choose n}\sum_{k=0}^n {n\choose k}\frac{\left(-\frac{1}{2}\right)_k}{\left(\frac{1}{2}-n\right)_k}(-x)^k\\
&={-\frac{1}{2}\choose n}\pFq{2}{1}{-\frac{1}{2},-n }{\frac{1}{2}-n}{x}
\end{align}
Using finally the identity
\be
{-\frac{1}{2}\choose n}=(-1)^n{2n\choose n}2^{-2n}\;,
\ee
we obtain the final expression in Eq. \eqref{surv_res}
\be
S_n^{+}(x=0;q)=\begin{cases}
&1\;,\;\;n=0\;,\\
&\displaystyle \frac{2^{-2n}}{1-q}{2n\choose n}\pFq{2}{1}{-\frac{1}{2},-n }{\frac{1}{2}-n}{2q-1}\;,\;\;n\geq 1\;.
\end{cases}
\ee

\subsection{Large $n$ limit for arbitrary initial position}

In the case of the exponential jump distribution, one can extract from Eqs. \eqref{ansatz} and \eqref{A_s_q} the behaviour of the survival probability for an arbitrary initial position. In particular, in the large $n$ limit and in the regime where $x=O(\sqrt{n})$, one obtains for $0\leq q<1$,
\be
\tilde S^{\pm}\left(x;s=1-\frac{a}{n};q\right)\approx \frac{n}{a}\left(1-e^{-\frac{x}{\sqrt{n}}\sqrt{2a(1-q)}}\right)
\ee
Using the Laplace inversion formula,
\be
{\cal L}_{a\to t}^{-1}\left(\frac{1}{a}\left(1-e^{-x\sqrt{a}}\right)\right)=\erf\left(\frac{x}{2\sqrt{t}}\right)\;,
\ee
it yields
\be
S_n^{\pm}\left(x;q\right)\approx \erf\left(\frac{x}{2\sqrt{n_{\rm eff}(q)}}\right)\;,
\ee
where $n_{\rm eff}(q)$ is given in Eq. \eqref{n_eff}. For $q=1/2$, we recover the well-known result for the survival probability of a random walk with finite variance (equal to $2$ here), which converges to that of the Brownian motion.

In the scaling regime where $n\to \infty$ and $q\to 1$ with $y=n(1-q)=O(1)$, the survival probability reads instead for $x=O(n)$,
\begin{align}
\tilde S^{+}\left(x;s=1-\frac{a}{n};q=1-\frac{y}{n}\right)&\approx \frac{n}{a}\left(1-e^{-\frac{x}{n}\sqrt{a(a+2y)}}\right)+\frac{n}{y}\left(\sqrt{\frac{a+2y}{a}}-1\right)e^{-\frac{x}{n}\sqrt{a(a+2y)}}\nn\\
\tilde S^{-}\left(x;s=1-\frac{a}{n};q=1-\frac{y}{n}\right)&\approx \frac{n}{a}\left(1-e^{-\frac{x}{n}\sqrt{a(a+2y)}}\right)
\end{align}
From this expression and using the Laplace inversion formulae
\begin{align}
{\cal L}_{a\to t}^{-1}\left(e^{-x\sqrt{a(a+2y)}}\right)=&\frac{y x e^{-y t}}{\sqrt{t^2-x^2}}\I_1(y\sqrt{t^2-x^2})+e^{-y t}\delta(t-x)\;,\\
{\cal L}_{a\to t}^{-1}\left(\sqrt{\frac{a+2y}{a}}e^{-x\sqrt{a(a+2y)}}\right)=&y e^{-y t}\left[\I_0(y\sqrt{t^2-x^2})+t\frac{\I_1(y\sqrt{t^2-x^2})}{\sqrt{t^2-x^2}}\right]\Theta(t-x)\nn\\
&+e^{-y t}\delta(t-x)\;,
\end{align}
it yields
\begin{align}
S_n^{\pm}\left(x;q\right)&\approx {\cal S}^{\pm}_{\rm RTP}\left(\frac{x}{n};n(1-q)\right)\;,\\
{\cal S}^{-}_{\rm RTP}(x;y)&=1-\Theta(1-x)e^{-y x}-\int_0^{1}d\tau \frac{y x \Theta(\tau-x)}{\sqrt{\tau-x^2}}e^{-y\tau}\I_1(y\sqrt{\tau^2-x^2})\\
{\cal S}^{+}_{\rm RTP}(x;y)&={\cal S}^{-}_{\rm RTP}(x;y)+\Theta(1-x)e^{-y}\left[\I_0(y\sqrt{1-x^2})+\sqrt{\frac{1-x}{1+x}}\I_1(y\sqrt{1-x^2})\right]\nn
\end{align}
One thus recovers the results for the survival probability for the RTP, starting from an arbitrary position $x\geq 0$ obtained in \cite{LDMS19,SK19}.

\section{Recovering the survival probability of RTP for $q=0$}

As previously mentioned, the positions of the random walk in the special case $q=0$ can be mapped exactly to the positions of a RTP at successive tumbling events. Note that the positions of the particle at these tumbling events are always local extrema of the trajectory. For a given trajectory, the RTP survives up to time $t$ if and only if all the positions at the tumbling events 
\be
0\leq t_1=\tau_1\leq t_2=\tau_1+\tau_2\leq\cdots\leq t_n=\sum_{k=1}^n \tau_k\leq t\;,
\ee 
and the final position $x(t)$ at time $t$ are positive. Note that for a general distribution, the distribution of the last step $x(t)-x_n$ can be quite different from $p(\eta)$. We consider here the exponential distribution of tumbling events $p(\tau)=\gamma e^{-\gamma \tau}$.
%
The survival probability at time $t$, starting from $x=0$ with a positive speed $+v_0$ is then
\begin{align}
S_+(t)&=\sum_{n\geq 0}\int_{0}^{\infty}d\tau_1\,\ldots\int_{0}^{\infty}d\tau_n\int_{0}^{\infty}d\tau_f \,\Theta_{n+1}(\tau_1,\cdots, \tau_n,\tau_f)\nn\\
&\times\prod_{k=1}^n p(\tau_i)p_f(\tau_f)\delta\left(\sum_{i=1}^{n}\tau_i+\tau_f-t\right)\;,
\end{align}
where the function $\Theta_{n}(y_1,\cdots, y_n)$ is defined in Eq. \eqref{def_theta}. In this expression, the distribution $p_f(\tau)=p(\tau)/\gamma=e^{-\gamma\tau}$. Taking the Laplace transform of this probability with respect to $t$, we obtain that
\begin{align}
\tilde S_+(s)&=\sum_{n\geq 0}\int_{0}^{\infty}d\tau_1 e^{-s \tau_1}\,\ldots\int_{0}^{\infty}d\tau_n e^{-s \tau_n}\int_{0}^{\infty}d\tau_f e^{-s \tau_f}\,\Theta_{n+1}(\tau_1,\cdots, \tau_n,\tau_f)\nn\\
&\times \prod_{k=1}^n p(\tau_i) p_f(\tau_f)=\sum_{n\geq 0}\left[\tilde p(s)\right]^n \tilde p_f(s)Q_{n+1}\;,\label{S_p_int}
\end{align}
where we have defined the Laplace transforms
\be
\tilde p(s)=\int_0^{\infty}p(\tau)e^{-s \tau}d\tau=\frac{\gamma}{\gamma+s}\;,\;\;\tilde p_f(s)=\int_0^{\infty}p_f(\tau)e^{-s \tau}d\tau=\frac{1}{\gamma+s}\;,
\ee
and $Q_{n+1}$ is defined here as
\be
Q_{n+1}=\int_{0}^{\infty}d\tau_1\ldots\int_{0}^{\infty}d\tau_n\int_{0}^{\infty}d\tau_f\,\Theta_{n+1}(\tau_1,\cdots, \tau_n,\tau_f)\prod_{k=1}^n \frac{p(\tau_i)e^{-s\tau_i}}{\tilde p(s)} \frac{p_f(\tau_f)e^{-s\tau_f}}{\tilde p_f(s)}\;.
\ee
Note that as the distributions $p(\tau)e^{-s\tau}/\tilde p(s)$ and $p_f(\tau)e^{-s\tau}/\tilde p_f(s)$ are both positive and normalised to unity in the interval $\tau>0$, they can both be interpreted as PDFs. In the particular case of the exponential distribution where $p_f(\tau)=p(\tau)/\gamma$, one simply has that $p_f(\tau)e^{-s\tau}/\tilde p_f(s)=p(\tau)e^{-s\tau}/\tilde p(s)$ and the $(n+1)$-fold integral  $Q_{n+1}$ can be interpreted as the universal survival probability of an alternating random walk given by $Q_{n+1}=S_{n+1}^+(x=0;q=0)$ in Eq. \eqref{S_p_q_0}. Using this result together with the inverse Laplace transform
\be
{\cal L}^{-1}_{s\to t}\left(\left[\tilde p(s)\right]^n \tilde p_f(s)\right)={\cal L}^{-1}_{s\to t}\left(\frac{\gamma^{n}}{(\gamma+s)^{n+1}}\right)=\frac{(\gamma t)^n}{n!}e^{-\gamma t}\;,
\ee
the probability $S_+(t)$ in Eq. \eqref{S_p_int} can be computed exactly as
\begin{align}
S_+(t)&=e^{-\gamma t}\sum_{p=0}^{\infty}\left[\frac{1}{(2p)!}{{2p}\choose{p}}\left(\frac{\gamma t}{2}\right)^{2p}+\frac{1}{2(2p+1)!}{{2(p+1)}\choose{p+1}}\left(\frac{\gamma t}{2}\right)^{2p+1}\right]\nn\\
&=e^{-\gamma t}\sum_{p=0}^{\infty}\left[\frac{1}{(p!)^2}\left(\frac{\gamma t}{2}\right)^{2p}+\frac{1}{p!(p+1)!}\left(\frac{\gamma t}{2}\right)^{2p+1}\right]\nn\\
&=e^{-\gamma t}\left(\I_0(\gamma t)+\I_1(\gamma t)\right)={\cal S}(\gamma t)\;,
\end{align}
recovering the result of \cite{LDMS19} (see also \cite{MDMS20,MDMS20_2} for an extension to the survival probability in arbitrary dimension $d\geq 1$). Here we used the Taylor expansion \cite{DLMF_I}
\be
I_p(x)=\sum_{n\geq 0}\frac{1}{n!(n+p)!}\left(\frac{x}{2}\right)^{2n+p}\;,\;\;p\in \mathbb{N}\;.
\ee

\end{document}